\documentclass[iop]{emulateapj}

\usepackage[colorlinks,urlcolor=blue,citecolor=blue,linkcolor=blue]{hyperref}
\usepackage{booktabs}
\usepackage{natbib,amsmath,textcomp}
\usepackage{pifont} 
\usepackage{color} 
\definecolor{grey}{rgb}{.7,.7,.7}
\slugcomment{DRAFT \today}
\shorttitle{PTF Star Galaxy Separation}
\shortauthors{Miller et al.}

\begin{document}

\title{Preparing for advanced LIGO: \\
A Star-Galaxy Separation Catalog for the Palomar Transient Factory}

\author{A.~A.~Miller\altaffilmark{1,2,3*}, M.~K.~Kulkarni\altaffilmark{2,4}, Y. Cao\altaffilmark{2}, R.~R.~Laher\altaffilmark{5}, F.~J.~Masci\altaffilmark{6}, \& J.~A.~Surace\altaffilmark{5} 
}

\altaffiltext{1}{Jet Propulsion Laboratory, California Institute of Technology, 4800 Oak Grove Drive, MS 169-506, Pasadena, CA 91109, USA}
\altaffiltext{2}{California Institute of Technology, Pasadena, CA 91125, USA}
\altaffiltext{3}{Hubble Fellow}
\altaffiltext{4}{University of California -- Berkeley, Berkeley, CA 94720, USA}
\altaffiltext{5}{Spitzer Science Center, California Institute of Technology, Pasadena, CA 91125, USA}
\altaffiltext{6}{Infrared Processing and Analysis Center, California Institute of Technology, Pasadena, CA 91125, USA}
\altaffiltext{*}{E-mail: {\tt amiller@astro.caltech.edu}}

\begin{abstract}

The search for fast optical transients, such as the expected electromagnetic counterparts to binary neutron star mergers, is riddled with false positives ranging from asteroids to stellar flares. While moving objects are readily rejected via image pairs separated by $\sim$1 hr, stellar flares represent a challenging foreground that significantly outnumber rapidly-evolving explosions. Identifying stellar sources close to and fainter than the transient detection limit can eliminate these false positives. Here, we present a method to reliably identify stars in deep co-adds of Palomar Transient Factory (PTF) imaging. Our machine-learning methodology utilizes the random forest (RF) algorithm, which is trained using $> 3\times{10}^6$ sources with Sloan Digital Sky Survey (SDSS) spectra. When evaluated on an independent test set, the PTF RF model outperforms the \texttt{SExtractor} star classifier by $\sim$4\%. For faint sources ($r'\ge{21}$ mag), which dominate the field population, the PTF RF model produces a $\sim$19\% improvement over \texttt{SExtractor}. To avoid false negatives in the PTF transient-candidate stream, we adopt a conservative stellar classification threshold, corresponding to a galaxy misclassification rate = 0.005. Ultimately, $\sim$$1.70\times{10}^8$ objects are included in our PTF point-source catalog, of which only $\sim$$10^6$ are expected to be galaxies. We demonstrate that the PTF RF catalog reveals transients that otherwise would have been missed. To leverage its superior image quality, we additionally create an SDSS point-source catalog, which is also tuned to have a galaxy misclassification rate = 0.005. These catalogs have been incorporated into the PTF real-time pipelines to automatically reject stellar sources as non-extragalactic transients.

\end{abstract}


\section{Introduction}\label{intro}

The classification or separation of stars vs.\ galaxies in astronomical images is an old problem with many important modern applications. At a very basic level, number counts of bright galaxies as a function magnitude show that the universe does not have a homogeneous ``Euclidean'' geometry \citep{Yasuda01}. More importantly, the accurate separation of stars and galaxies in faint samples significantly improves our ability to (i) measure galaxy-galaxy correlation functions (e.g., \citealt{Ross11, Ho15}, (ii) map the signature of baryon acoustic oscillations \citep{Anderson14}, (iii) search for dwarf galaxies by looking for stellar overdensities (e..g, \citealt{Belokurov07}), (iv) detect the weak-lensing signal from cosmic shear (\citealt{Soumagnac15} and references therein), and (v) trace structure in the Milky Way halo (e.g., \citealt{Belokurov06, Juric08}), among other things. 

The array of scientific problems dependent upon star-galaxy separation is disparate, meaning the construction of any such catalog should be application specific. For time-domain surveys aiming to identify transients, a reliable star-galaxy catalog immediately informs researchers of the galactic or extragalactic origin of newly discovered candidates. 

The Palomar Transient Factory (PTF; \citealt{rau09,law09}) is a dedicated survey of the variable sky utilizing the CFH12K mosaic camera on the Palomar 48-inch telescope (P48). The initial phase of this experiment ended in 2012, while the current iteration, the intermediate Palomar Transient Factory (iPTF; \citealt{Kulkarni13}) started in 2013. The next generation Palomar time-domain survey, the Zwicky Transient Facility (ZTF; \citealt{Kulkarni12}), will begin in 2017. ZTF will upgrade the camera on the P48 and feature improved electronics and a $\sim$47 $\deg^2$ field of view (FOV), which is a factor of $\sim$7 increase over the PTF FOV. 

A primary motivation for both PTF and ZTF is the search for fast ($\lesssim 24$ hr) transients, a rare class of explosive events expected to include ``kilonovae'', the result of binary neutron star (BNS) mergers (e.g., \citealt{Kasen15}). BNS mergers are thought to be the most promising electromagnetic counterparts to gravitational wave (GW) sources (e.g., \citealt{Metzger12,Nissanke13}). Now that we are firmly in the age of GW detections \citep{Abbott16}, the search for electromagnetic counterparts is both highly exciting and extremely pressing. As surveys identify fast-transient candidates, including GW counterparts, they contend with significant foreground contamination in the form of stellar flares (e.g., \citealt{Kulkarni06,Berger12}). The systematic removal of faint stars from extragalactic candidate lists can fully alleviate this problem by removing false positives from consideration for expensive follow-up resources. Indeed, while searching for an optical counterpart to GW150914, a (now outdated) PTF star catalog rejected $\sim$40\% of the viable transient candidates \citep{Kasliwal16}.

\begin{figure*}
\centerline{\includegraphics[width=7.2in]{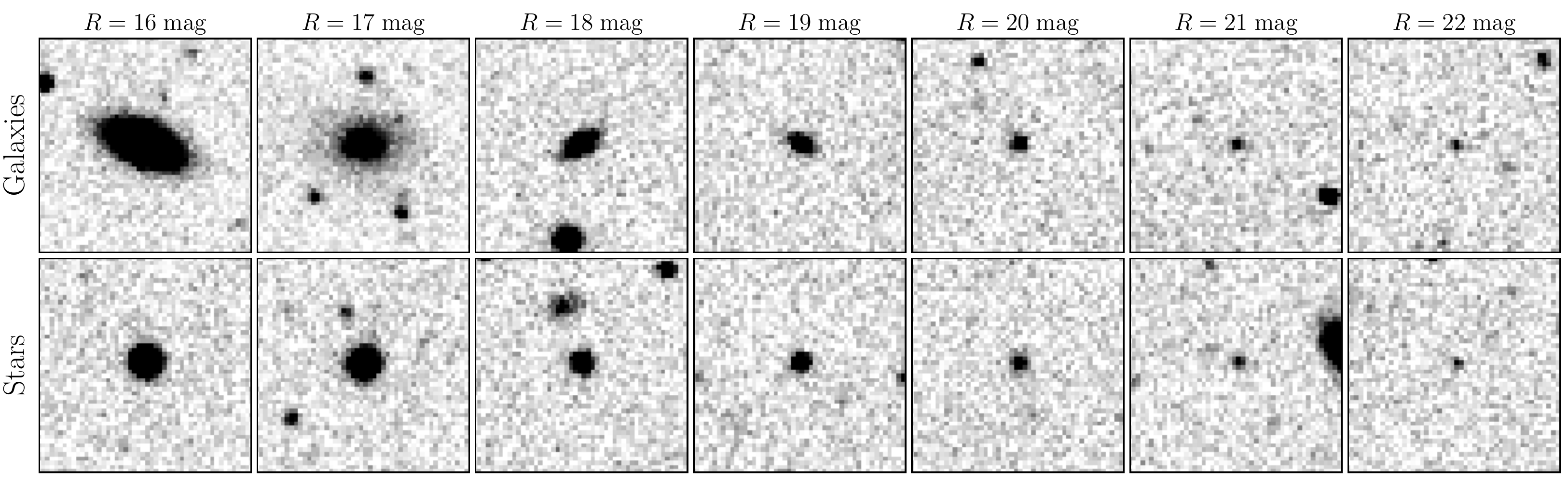}}
\caption[]{Postage stamps showing typical stars and galaxies in PTF reference images as a function of magnitude. The images show that stars and galaxies can easily be separated by eye down to R $\approx$ 19 mag, while for fainter sources the two are virtually indistinguishable. Postage stamps are $50 \times 50$ pixels, centered on the source of interest, with north up and east to the left. Source classifications are from SDSS spectra. Each stamp is from a reference coadd of 5 individual PTF images, the shallowest reference images produced by PTF, yielding an effective exposure time of 300 s. 
}
\label{fig:sG_rogues}
\end{figure*}

PTF employs sophisticated software solutions to rapidly process new observations, perform image subtraction, and identify transient candidates (\citealt{Cao16}, ; \citealt{Masci16}). These candidates are then confirmed or rejected as bonafide astrophysical variations by machine-learning software (e.g., \citealt{Bloom12,Brink13,Rebbapragada15}). At this stage human vetting of the candidates identifies those that merit additional follow-up observations. Within the Sloan Digital Sky Survey (SDSS; \citealt{York00}) imaging footprint stars and galaxies can be separated with high fidelity to a faintness of $\sim$22 mag, by comparing the point-spread-function (PSF) magnitude to the best-fit model magnitude.\footnote{See \url{http://www.sdss.org/dr12/algorithms/classify/\#photo_class} for further details.} However, SDSS only overlaps $\sim$half of the full PTF imaging footprint, and faint objects cannot reliably be classified as stars or galaxies via visual inspection in PTF images, as illustrated in Figure~\ref{fig:sG_rogues}.

It naturally follows that the development of a star-galaxy-separation model for PTF would improve our ability to reject false positives in our search for fast transients and GW counterparts. The optimal star-galaxy catalog for fast-transient surveys would identify as many stars as possible (true positives), while minimizing the number of galaxies misclassified as stars (false positives). Striking the proper balance between these two objectives is challenging: an overly conservative selection of stars will result in many transient candidates with Galactic origin, while an overly aggressive selection will lead to many galaxies being excluded from the search. The intrinsic rarity of fast transients and GW counterparts means the latter situation, which could result in a GW counterpart being missed entirely, is especially undesirable. Machine-learning algorithms offer an attractive solution to this problem as they enable a precise tuning of the classification decision threshold to balance the number of true positives and false positives. 

Supervised machine-learning algorithms construct a model to map \textit{features}, measured properties of the sources, to \textit{labels}, such as a classification or physical property.\footnote{For a more detailed primer on machine learning, we refer the reader to \citet{Hastie09}.} The model is constructed using a training set, and its performance is evaluated using a test set. The training set and test set are independent subsets of the data with (spectroscopic) labels that we adopt as ground truth. Machine-learning models are very flexible, capable of capturing complex nonlinear behavior in the multidimensional feature space. In many cases they provide fast, automated classifications for new data. Previously, machine-learning models utilizing decision trees have been used to successfully classify stars and galaxies in SDSS imaging data \citep{Ball06,Vasconcellos11}. 

Here, we construct an ensemble of decision trees model, trained with spectroscopic classifications from SDSS, to separate stars and galaxies in PTF images. We describe our procedure to curate an appropriate training set and the steps utilized to optimize the performance of the algorithm. Most importantly, we compare the performance of our model to that of \texttt{SExtractor}, which currently provides the best discriminant between stars and galaxies in PTF images outside the SDSS photometric footprint. We define conservative selection criteria for stellar classification, and apply the final optimized model to all $> 5 \times 10^8$ sources in the PTF photometric catalog. This catalog has been ingested by the appropriate PTF pipelines, and is currently used to reject false positives in the search for new transients.

\section{Training the Model with SDSS Spectroscopic Targets}\label{sec:training_set}

An important and essential first step in the construction of a supervised machine-learning model is the curation of the training set and test set. The data-driven nature of supervised machine learning means that special consideration must be taken to avoid potential biases in the training set. The final model predictions for the full data set will reflect, and likely preserve, any biases in the training set. For the PTF star-galaxy catalog, features are extracted from PTF reference images (deep coadds) using \texttt{SExtractor} \citep{bertin96}, and labels are provided by SDSS spectroscopic observations. 

\subsection{SDSS Training Labels}

To facilitate the search for transient sources, the PTF imaging pipeline produces reference images \citep{Laher14}, deep coadds of $\ge$5 individual 60 s exposures. PTF reference images are significantly deeper and offer superior image quality to individual exposures. We employ only the $R$-band detections for the model because there are significant gaps in the sky coverage for the other PTF filters. To train our model, we consider all photometric detections from PTF $R$-band reference images. Using all $R$-band references available as of 2016 July 22 UT, there are 548,687,903 sources detected by \texttt{SExtractor}. PTF employs a grid of overlapping pointings, thus, some of those $\sim$550 million detections represent duplicates of the same astrophysical source.

To identify which photometric detections are suitable to train the machine-learning model, we adopt the labels from SDSS spectroscopic classifications as ``ground truth.'' Optical spectra taken as part of the original SDSS survey and the Baryon Oscillation Spectroscopic Survey (BOSS; \citealt{Dawson13}) were automatically classified as belonging to one of three classes: stars, galaxies, and quasi-stellar objects (QSOs). Using PTF imaging data, we hope to separate resolved (galaxies) and unresolved (stars, QSOs) sources, which for simplicity will be hereafter referred to as galaxies and stars, respectively. 

Using the spatial crossmatch tool available via SDSS \texttt{CasJobs}, all PTF photometric sources coincident within 1$\arcsec$ of an SDSS spectroscopic source are selected as potential training objects. In total, there are 3,193,349 matches between PTF and the SDSS spectroscopic catalog. To prevent over-fitting, these sources were split roughly 60-40 into independent training and test sets. The training and test sets are used to optimize the model and evaluate its accuracy, defined as the fraction of sources that are correctly classified, respectively. As previously mentioned, there are photometric duplicates in the PTF reference image catalogs. In addition to this, SDSS obtained spectra of some sources more than once. Thus, a random 60-40 split of the $\sim$3 million training objects would not ensure independence between the training and test sets. To prevent sources from being assigned to both sets, we randomly select 60\% of the unique \texttt{objid}, the SDSS photometric identification key, and assign all sources with matching \texttt{objid} to the training set. All remaining sources are assigned to the test set. Following this procedure, the training set includes 1,919,088 sources while the test set has 1,274,261 sources.

Qualitatively, the distribution of the number of coadds, $N_\mathrm{coadd}$, in the reference image on which a source is detected is similar for the full $\sim$550 million PTF source catalog and the $\sim$3 million training sources. For both the full catalog and the training set a plurality of sources have $N_\mathrm{coadd} = 5$, 46\% and 38\%, respectively. Both distributions exhibit strong positive skew, with a secondary peak at $N_\mathrm{coadd} = 50$, the maximum number of coadds. Ultimately, the training set is more biased towards deep images with 11\% of sources having $N_\mathrm{coadd} \ge 30$, while the same is true for only 5\% of sources in the full catalog. Nevertheless, we do not expect these differences to produce significant biases in the final star-galaxy predictions because the overall distributions are similar, and the training set is slightly deeper and less noisy. 

\subsection{\texttt{SExtractor} Photometric Features}\label{sec:SEfeats}

The PTF reference-image pipeline utilizes \texttt{SExtractor} to measure 96 photometric properties per source. Relevant features for classifying sources as either stars or galaxies include: elongation, full-width half-max (FWHM), best-fit Petrosian radius, etc. Several properties measured by \texttt{SExtractor} are contextual, such as the \texttt{X} and \texttt{Y} position of the source photocenter on the CCD, and we exclude these from the machine-learning model. Furthermore, we normalize all \texttt{SExtractor} shape measurements by the average seeing in a given image\footnote{For PTF the seeing is determined from a trimmed mean of the \texttt{FWHM\_IMAGE} parameter measured by \texttt{SExtractor} \citep{Laher14}.} and all flux measurements by the flux in a circular aperture with 2 pixel diameter. The former accounts for the variable observing conditions for different references, while the latter helps to remove biases due to the brightness distribution of SDSS spectroscopic targets (see \S\ref{sec:results}).\footnote{While contectual information, such as brightness or galactic latitude, could in principle help data-driven classification, in many cases contextual features propagate biases in target selection to the final model (see e.g., \citealt{Richards12a}). Hence, we exclude positional coordinates and normalize brightness measurements for our final model. 
} 

\begin{deluxetable*}{ll}
\tabletypesize{\small}
\tablecolumns{2}
\tablecaption{PTF \texttt{SExtractor} Features Excluded from the Model\label{tbl:excluded_feats}}
\tablehead{\colhead{Name} & \colhead{Description} }
\startdata
\texttt{NUMBER} & 
Identification number of object \\

\texttt{X\_IMAGE}, \texttt{Y\_IMAGE}  & 
Pixel position of object centroid. \\

\texttt{XWIN\_IMAGE}, \texttt{YWIN\_IMAGE} & 
Pixel position of object centroid, windowed measurement. \\

\texttt{X\_WORLD}, \texttt{Y\_WORLD}  & 
RA and Dec coordinates of object centroid. \\

\texttt{XPEAK\_IMAGE}, \texttt{YPEAK\_IMAGE} & 
Pixel position with peak object intensity \\

\texttt{ALPHAWIN\_J2000} & 
Right ascension of object barycenter (J2000) \\

\texttt{DELTAWIN\_J2000} & 
Declination of object barycenter (J2000) \\

\texttt{THETAWIN\_J2000} & 
Object position angle (east of north) (J2000) \\ 

\texttt{CLASS\_STAR} &   
\texttt{SExtractor} stellarity index between 0-1. 1 = star
\\

\texttt{MAG\_ISO} & 
Isophotal Magnitude measurement \\
\color{grey}\texttt{MAGERR\_ISO} \ &  \\

\texttt{MAG\_ISOCOR} &  
Corrected isophotal magnitude \\
\color{grey} \texttt{MAGERR\_ISOCOR}  & \\

\texttt{MAG\_BEST} & 
Best of \texttt{MAG\_AUTO} and \texttt{MAG\_ISOCOR} \\
\color{grey}  \texttt{MAGERR\_BEST}   & \\

\texttt{MU\_THRESHOLD} & 
Surface brightness detection threshold above background \\

\texttt{BACKGROUND} & 
Background at object centroid position \\

\texttt{THRESHOLD} & 
Detection threshold above background \\

\texttt{ISOAREA\_WORLD} &  
Isophotal area above threshold \\

\texttt{ISOAREAF\_WORLD}   & Isophotal area (filtered) above threshold (degrees) \\

\texttt{ISO0}, \texttt{ISO1}, \texttt{ISO2}, \texttt{ISO3}, & \\
\texttt{ISO4}, \texttt{ISO5}, \texttt{ISO6}, \texttt{ISO7}   & 
Isophotal area at level \emph{n} \\

\texttt{FLUX\_BEST} & 
Best of \texttt{FLUX\_AUTO} and \texttt{FLUX\_ISOCOR} \\
\color{grey} \texttt{FLUXERR\_BEST}  & \\

\texttt{FLUX\_AUTO} & 
Flux within a Kron-like elliptical aperture \\
\color{grey}  \texttt{FLUXERR\_AUTO}   & \\

\texttt{FLUX\_ISO} & 
Isophotal Flux \\
\color{grey}  \texttt{FLUXERR\_ISO}   & 

\enddata
\tablecomments{
Feature names shown in grey, all of which include \texttt{ERR}, indicate uncertainty measurements for the \textbf{immediately} preceding feature. The use of isophotal magnitude measurements has been deprecated in \texttt{SExtractor}, therefore, we exclude these features (\texttt{MAG\_ISO}, \texttt{MAG\_ISOCOR}, \texttt{MAG\_BEST}) from the final model. 
 }
\end{deluxetable*}

The list of \texttt{SExtractor} measurements that are excluded from the machine-learning model are listed in Table~\ref{tbl:excluded_feats}.\footnote{For a full description of all \texttt{SExtractor} features see the \href{https://www.astromatic.net/pubsvn/software/sextractor/trunk/doc/sextractor.pdf}{documentation} or Dr. Benne Holwerda's excellent \href{http://astroa.physics.metu.edu.tr/MANUALS/sextractor/Guide2source_extractor.pdf}{Guide to SExtractor}.} The majority of the features in this table are uniformative. One major exception is \texttt{CLASS\_STAR}, which is a neural-network based source classification ranging from 0 to 1. Sources with \texttt{CLASS\_STAR} $\approx$ 1 are considered star-like, while sources with \texttt{CLASS\_STAR} $\approx$ 0 are non-star-like (galaxies, but also cosmic rays, etc.). Outside the SDSS photometric footprint, \texttt{CLASS\_STAR} represents the best model for separating stars and galaxies in PTF data. Thus, we exclude \texttt{CLASS\_STAR} from the model so that we may compare our final classifications against those made by \texttt{SExtractor}.

\begin{figure*}
\centerline{\includegraphics[width=7in]{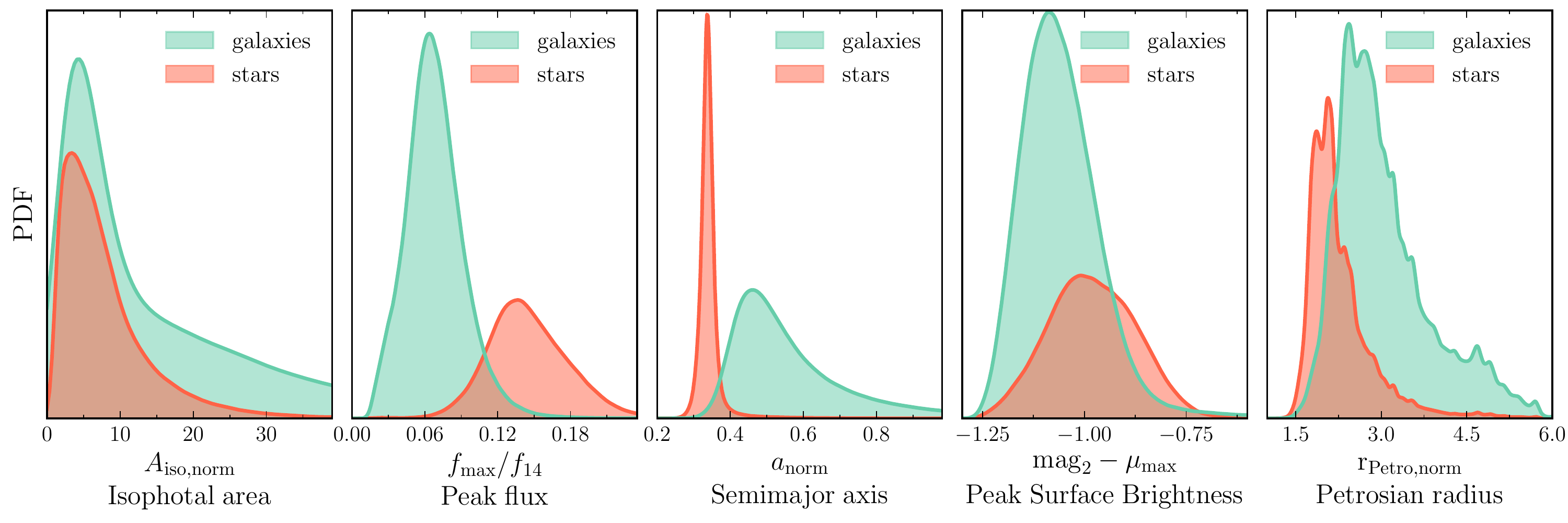}}
\caption[]{KDE of the PDF for of select model features for all PTF sources with spectroscopic observations. The area under the curves has been scaled realtive to the total number of stars and galaxies in the spectroscopic set (galaxies outnumber stars by a factor of $\sim$2:1). Distributions are for the normalized features (see the text and Table~\ref{tbl:included_feats} for further details). While no single feature separates stars and galaxies significantly better than \texttt{CLASS\_STAR}, the differing PDFs for the two populations suggests a non-parametric method may produce a significant improvement over the \texttt{SExtractor} stellarity estimate.}
\label{fig:sG_feats}
\end{figure*}

Features provided to the machine-learning model are listed in Table~\ref{tbl:included_feats}. As previously noted, shape parameters are normalized by the average seeing in a reference image, while all magnitude measurements are normalized relative to the magnitude measured in a 2 pixel diameter circular aperture. We exclude the uncertainties on the shape and brightness measurements from the model as these primarily reflect the depth of the reference image, which varies considerably over the dataset given that some coadds include 5 images while others include 50. Following normalization, we supply the machine-learning model with 43 features. A kernel density estimate (KDE)\footnote{All KDEs presented in this paper adopt a Gaussian kernel and Scott's rule to determine the kernel bandwidth \citep{Scott92}.} of the probability distribution function (PDF) of 5 uncorrelated features is shown in Figure~\ref{fig:sG_feats}.

\begin{deluxetable*}{lll}
\tabletypesize{\small}
\tablecolumns{3}
\tablecaption{PTF \texttt{SExtractor} Features Included in the Model\label{tbl:included_feats}}
\tablehead{\colhead{Name} & \colhead{Description} & \colhead{Normalization Factor}}
\startdata
\texttt{X2\_IMAGE}  & 
Second order moment of object, along $x$-axis. & 
$1/\mathrm{seeing}^2$ \\

\texttt{Y2\_IMAGE}  & 
Second order moment of object, along $y$-axis. & 
$1/\mathrm{seeing}^2$ \\

\texttt{X2WIN\_IMAGE}  & 
Second order moment of object, windowed measurement. & 
$1/\mathrm{seeing}^2$ \\
\color{grey} \texttt{ERRX2WIN\_IMAGE} & &\\

\texttt{Y2WIN\_IMAGE}  & 
Second order moment of object, windowed measurement. & 
$1/\mathrm{seeing}^2$ \\
\color{grey} \texttt{ERRY2WIN\_IMAGE} & &\\

\texttt{XY\_IMAGE} & 
Covariance of position between x and y. & 
$1/\mathrm{seeing}^2$ \\

\texttt{XYWIN\_IMAGE} &
Covariance of position between x and y, windowed measurement. &
$1/\mathrm{seeing}^2$ \\
\color{grey}\texttt{ERRXYWIN\_IMAGE} & & \\

\texttt{AWIN\_WORLD} & 
Object profile rms along the major axis, windowed measurement. &
$1/\mathrm{seeing}$ \\
\color{grey}\texttt{ERRAWIN\_IMAGE} & & \\

\texttt{BWIN\_WORLD} &
Object profile rms along the minor axis, windowed measurement. &
$1/\mathrm{seeing}$ \\
\color{grey}\texttt{ERRBWIN\_IMAGE} & & \\

\texttt{MAG\_APER}$_{2}$ &  
Magnitude in a 2 pixel diameter circular aperture centered on the object. &
See table notes.
\\
\color{grey}  \texttt{MAGERR\_APER}$_{2}$  & & \\

\texttt{MAG\_APER}$_{4}$ &  
Magnitude in a 4 pixel diameter circular aperture centered on the object. &
\texttt{MAG\_APER}$_{2}$
\\
\color{grey}  \texttt{MAGERR\_APER}$_{4}$  & & \\

\texttt{MAG\_APER}$_{5}$ &  
Magnitude in a 5 pixel diameter circular aperture centered on the object. &
\texttt{MAG\_APER}$_{2}$
\\
\color{grey}  \texttt{MAGERR\_APER}$_{5}$  & & \\

\texttt{MAG\_APER}$_{8}$ &  
Magnitude in a 8 pixel diameter circular aperture centered on the object. &
\texttt{MAG\_APER}$_{2}$
\\
\color{grey}  \texttt{MAGERR\_APER}$_{8}$  & & \\

\texttt{MAG\_APER}$_{10}$ &  
Magnitude in a 10 pixel diameter circular aperture centered on the object. &
\texttt{MAG\_APER}$_{2}$
\\
\color{grey}  \texttt{MAGERR\_APER}$_{10}$  & & \\

\texttt{MAG\_APER}$_{14}$ &  
Magnitude in a 14 pixel diameter circular aperture centered on the object. &
\texttt{MAG\_APER}$_{2}$
\\
\color{grey}  \texttt{MAGERR\_APER}$_{14}$  & & \\

\texttt{MAG\_AUTO} &  
Kron-like elliptical aperture magnitude. &
\texttt{MAG\_APER}$_{2}$
\\
\color{grey} \texttt{MAGERR\_AUTO} & &  \\

\texttt{MAG\_PETRO} &   
Petrosian-like elliptical aperture magnitude. & 
\texttt{MAG\_APER}$_{2}$
\\
\color{grey} \texttt{MAGERR\_PETRO}  & & \\

\texttt{MU\_MAX} &   
Peak surface brightness above background. &
\texttt{MAG\_APER}$_{2}$ 
\\

\texttt{THETA\_IMAGE} &  
Position angle of object, counter clockwise. & 
\\

\texttt{THETAWIN\_IMAGE}   & 
Position angle of object, counter clockwise, windowed measurement. & 
\\
\color{grey} \texttt{ERRTHETAWIN\_IMAGE} & &\\

\texttt{THETAWIN\_WORLD} & 
Position angle of object, counter clockwise, world coordinates.
\\

\texttt{ELONGATION} & 
\texttt{A\_IMAGE/B\_IMAGE} & 
\\

\texttt{FWHM\_IMAGE} &  
Full-Width Half Max of object, assuming Gaussian core. & 
$1/\mathrm{seeing}$ 
\\

\texttt{KRON\_RADIUS} &  
Kron radius of object. & 
$1/\mathrm{seeing}$ 
\\

\texttt{PETRO\_RADIUS}   & 
Petrosian radius of object. & 
$1/\mathrm{seeing}$ 
\\

\texttt{ISOAREAF\_IMAGE} & 
Isophotal area (filtered) above threshold. &
$1/\mathrm{seeing}^2$
\\

\texttt{FLUX\_APER}$_{2}$ &  
Flux in a 2 pixel diameter circular aperture centered on the object. &
1/\texttt{FLUX\_MAX}
\\
\color{grey}  \texttt{FLUXERR\_APER}$_{2}$  & & \\

\texttt{FLUX\_APER}$_{4}$ &  
Flux in a 4 pixel diameter circular aperture centered on the object. &
1/\texttt{FLUX\_MAX}
\\
\color{grey}  \texttt{FLUXERR\_APER}$_{4}$  & & \\

\texttt{FLUX\_APER}$_{5}$ &  
Flux in a 5 pixel diameter circular aperture centered on the object. &
1/\texttt{FLUX\_MAX}
\\
\color{grey}  \texttt{FLUXERR\_APER}$_{5}$  & & \\

\texttt{FLUX\_APER}$_{8}$ &  
Flux in a 8 pixel diameter circular aperture centered on the object. &
1/\texttt{FLUX\_MAX}
\\
\color{grey}  \texttt{FLUXERR\_APER}$_{8}$  & & \\

\texttt{FLUX\_APER}$_{10}$ &  
Flux in a 10 pixel diameter circular aperture centered on the object. &
1/\texttt{FLUX\_MAX}
\\
\color{grey}  \texttt{FLUXERR\_APER}$_{10}$  & & \\

\texttt{FLUX\_APER}$_{14}$ &  
Flux in a 14 pixel diameter circular aperture centered on the object. &
1/\texttt{FLUX\_MAX}
\\
\color{grey}  \texttt{FLUXERR\_APER}$_{14}$  & & \\

\texttt{FLUX\_RADIUS}$_{25}$ &   
Radius enclosing 25\% of the object flux. &
$1/\mathrm{seeing}$\\

\texttt{FLUX\_RADIUS}$_{50}$ &   
Radius enclosing 50\% of the object flux. &
$1/\mathrm{seeing}$\\

\texttt{FLUX\_RADIUS}$_{85}$ &   
Radius enclosing 85\% of the object flux. &
$1/\mathrm{seeing}$\\

\texttt{FLUX\_RADIUS}$_{95}$ &   
Radius enclosing 95\% of the object flux. &
$1/\mathrm{seeing}$\\

\texttt{FLUX\_RADIUS}$_{99}$ &   
Radius enclosing 99\% of the object flux. &
$1/\mathrm{seeing}$\\

\texttt{FLUX\_MAX}   &
Flux of brightest pixel within the object. &
See table notes. \\

\texttt{FLAGS} & 
Source Extractor Flags, coded in bitmask. &

\enddata
\tablecomments{
Feature names shown in grey, all of which include \texttt{ERR}, indicate uncertainty measurements for the immedidately preceding feature. These measurements of the uncertainty are not used by the model. Normalization factors are multiplicative, aside from mag measurements, and are applied, as needed, prior to running the model. Magnitudes are logarithmic, thus all mag measurements are normalized via a difference (e.g., the Petrosian mag is represented as \texttt{MAG\_APER}$_{2}$ - \texttt{MAG\_PETRO}). In practice, the inverse of the aperture flux is used (e.g., the flux in a 5 pixel aperture is represented as \texttt{FLUX\_MAX}/\texttt{FLUX\_APER}$_{5}$). Each of the 8 \texttt{SExtractor} \texttt{FLAGS} are treated as a binary feature for the classifier. \texttt{MAG\_APER}$_{2}$ and \texttt{FLUX\_MAX} are not represented in the final model as they are otherwise captured as normalization factors. In sum, 43 features are included in the machine-learning model.
}
\end{deluxetable*}

\subsection{Removal of Photometric Blends}\label{sec:systematics}

During the final stages of model construction, we noticed an unusual systematic whereby a large fraction of stars with $r' \approx 20.5$ mag were erroneously classified as galaxies (see \S\ref{sec:results} and Figure~\ref{fig:stars_galaxies} for further details). Manual inspection of several of these sources revealed that they were red stars blended with fainter sources. The SDSS spectroscopic survey was intentionally biased towards observing luminous red galaxies (LRGs, see e.g., \citealt{Eisenstein01}), and thus faint red stars that are photometrically blended would satisfy the general LRG selection criteria of being red, faint, and extended. 

We found that these sources could be readily identified using both the \texttt{class} and \texttt{sourceType} columns in the \texttt{specObjAll} table of the DR12 SDSS database \citep{Alam15}. These columns elucidate the spectroscopic class (\texttt{GALAXY}, \texttt{QSO}, or \texttt{STAR}) of the target and the reason the source was targeted, respectively. Our ultimate goal with this classification catalog is to develop a pristine list of point-sources. In other words, we are willing to accept stars being classified as galaxies if those stars are blended with other sources such that their photometric appearance resembles galaxies. While such an approach would be disadvantageous to galaxy clustering studies, it is ideal for the search for transients. Thus, we exclude spectroscopic stars targeted as galaxies from the training set. Similarly, we exclude spectroscopic galaxies targeted as either stars or QSOs and spectroscopic QSOs targeted as galaxies. We additionally exclude all spectroscopic QSOs with redshift $z < 1$, many of which have detectable host galaxies in addition to their active galactic nuclei. Finally, we exclude a small number (34,437) of emission-line galaxy (ELG) and Sloan Extended Quasar, ELG, and LRG Survey (SEQUELS) targets, which consistently have spectroscopic classes that do not match their target class. 

\begin{deluxetable*}{lrrrrrr}
\tabletypesize{\small}
\tablecolumns{7}
\tablecaption{SDSS Spectroscopic Targets Excluded from the Training Set\label{tbl:excluded_spec}}
\tablehead{ \colhead{\texttt{class}} & \multicolumn{2}{c}{\texttt{STAR}} & \multicolumn{2}{c}{\texttt{QSO}} & \multicolumn{2}{c}{\texttt{GALAXY}} }
\startdata
\texttt{sourceType} & \texttt{LRG} & (16444) & 
\texttt{SEQUELS\_TARGET} & (16511) & 
\texttt{QSO} & (29106)\\
 & \texttt{GALAXY} & (11142) & 
 \texttt{LRG} & (16150) & 
 \texttt{SEQUELS\_TARGET} & (12210) \\
 & \texttt{HIZ\_LRG} & (169) & 
 \texttt{GALAXY} & (3293) & 
 \texttt{SEQUELS\_ELG} & (1545) \\
 &  &  & 
 \texttt{SN\_GAL1} & (1297) & 
 \texttt{FAINT\_ELG} & (1071) \\
 &  &  & 
 \texttt{SEQUELS\_ELG} & (528) & 
 \texttt{SEQUELS\_ELG\_LOWP} & (986) \\
 &  &  & 
 \texttt{ELG} & (525) & 
 \texttt{ELG} & (887) \\
 &  &  & 
 \texttt{SEQUELS\_ELG\_LOWP} & (174) & 
 \texttt{STAR} & (812) \\
 &  &  & 
 &  & 
 \texttt{WISE\_BOSS\_QSO} & (694) \\
 &  &  & 
 &  & 
 \texttt{QSO\_EBOSS\_W3\_ADM} & (548) \\
 &  &  & 
 &  & 
 \texttt{QSO\_WISE\_FULL\_SKY} & (371) \\
 &  &  & 
 &  & 
 \texttt{QSO\_VAR\_SDSS} & (362) \\
 &  &  & 
 &  & 
 \texttt{QSO\_VAR\_LF} & (324) \\
 &  &  & 
 &  & 
 \texttt{QSO\_WISE\_SUPP} & (269) \\
 &  &  & 
 &  & 
 \texttt{SERENDIPITY\_BLUE} & (262) \\
 &  &  & 
 &  & 
 \texttt{SERENDIPITY\_DISTANT} & (237) \\
 &  &  & 
 &  & 
 \texttt{QSO\_GRI} & (227) \\
 &  &  & 
 &  & 
 \texttt{QSO\_DEEP} & (147) \\
 &  &  & 
 &  & 
 \texttt{STD} & (128) \\
\hline
total & & 27755 & & 116568\tablenotemark{a} & & 50186

\enddata
\tablecomments{
For each spectroscopic class (\texttt{class}) the corresponding number of sources from each \texttt{sourceType} that are removed from the training and validation set \textit{combined} are shown in parentheses. }
\tablenotetext{a}{In addition to the 38,478 spectroscopic QSOs detailed above, an additional 78,090 spectroscopic QSOs with $z < 1$ are removed from the training and validation sets.
}
\end{deluxetable*}

The full list of SDSS spectroscopic class and target type combinations that were excluded from the training and test sets is summarized in Table~\ref{tbl:excluded_spec}. In total, these exclusions remove 194,509 sources from the 3,193,349 PTF sources with SDSS spectra. The final training and test sets, which we hereafter refer to as \textit{photometrically clean}, include 1,802,357 and 1,196,483 sources, respectively. 

\section{Machine-Learning Model Construction}\label{sec:model}

\subsection{The Random Forest Algorithm}\label{sec:RFalgorithm}

Random forest (RF) methods utilize the aggregation of multiple decision trees to assign a final classification or regression value to newly observed sources \citep{Breiman01}. RF makes use of bagging (see \citealt{Breiman96}), wherein bootstrap samples of the training set are used to construct each of the $N_{\rm tree}$ total trees in the forest. As each tree in the forest is constructed, only a random subset of $m_{\rm try}$ features is selected from the full feature set as a potential splitting criterion at each node of the tree. The use of bagging and $m_{\rm try}$ random features reduces the variance of the final model  relative to single decision-tree models, providing low-bias, low-variance predictions. The final RF predictions are determined by averaging the predictions for a new source from each of the $N_{\rm tree}$ individual trees. Furthermore, the RF algorithm is fast, each of the trees can be constructed independently and thus in parallel, and relatively easy to interpret. RF models have recently become highly popular as an application for astronomical data sets due to their relative insensitivity to noisy or meaningless features (e.g., \citealt{Brink13, Miller15}), and their invariant response to even highly non-gaussian feature distributions (e.g., \citealt{Richards11,Dubath11}). Due to its flexibility and speed, we adopt RF for this study, in particular, we utilize the \texttt{Python scikit-learn}\footnote{\url{http://scikit-learn.org/stable/}} implementation of the algorithm \citep{Pedregosa11}.

\subsection{Imputation for Missing Features}

An initial challenge for the classification model is that \texttt{SExtractor} does not always produce finite measurements for the features listed in \S\ref{sec:SEfeats}. For a small number of sources \texttt{BWIN\_WORLD} is reported as \texttt{NaN}, while a slightly larger number of sources have \texttt{APER\_FLUX} measurements of 0, which results in a normalized feature value of infinity (see Table~\ref{tbl:included_feats}). In the training set, a single source has a bad \texttt{BWIN\_WORLD} measurement while 48, 47, 47, 46, 45, and 44 sources have zero-valued aperture flux measurements from the smallest to the largest aperture, respectively. In the full, 548,687,903 source PTF reference catalog 731 sources have bad \texttt{BWIN\_WORLD} measurements, while 11958, 11421, 11160, 10687, 10595, and 10474 sources have zero-valued aperture flux measurements from the smallest to the largest aperture, respectively. 

There are three potential solutions to deal with missing features: (i) exclude any features with missing data from the model entirely, (ii) exclude the sources with missing features from the model training and final predictions, or (iii) develop a method to estimate the values of the missing features. The first two options are non-desirable as they remove information from the model and prevent predictions for some sources, respectively. The third option is most attractive as it does not exclude any valuable information.

We test two methods of imputation to estimate the missing values in the feature set. The first is simple: replace all missing values with the median value of the feature in the training set. The second is more complex: use RF regression to perform a nonparametric estimate of the missing features using the features with no missing values. In particular, we use RF regression with $N_\mathrm{tree} = 100$ and fully grown trees to estimate the missing values. \citet{Stekhoven12} have shown that this nonparametric method outperforms several other common methods for imputation. 

\begin{deluxetable*}{lccccccc}
\tabletypesize{\small}
\tablecolumns{8}
\tablecaption{Imputation Results\label{tbl:impute}}
\tablehead{ 
\colhead{ } & \colhead{\texttt{BWIN\_WORLD}} &
\colhead{\texttt{FLUX\_APER}$_{2}$} & \colhead{\texttt{FLUX\_APER}$_{4}$} &
\colhead{\texttt{FLUX\_APER}$_{5}$} & \colhead{\texttt{FLUX\_APER}$_{8}$} &
\colhead{\texttt{FLUX\_APER}$_{10}$} & \colhead{\texttt{FLUX\_APER}$_{14}$}
}
\startdata


\textbf{RF} & \textbf{0.028} & \textbf{6411.8} & \textbf{1589.8} & \textbf{1012.4} & \textbf{13.2} & \textbf{8.6} & \textbf{4.7} \\
\textbf{median} & \textbf{0.183 }& \textbf{6400.2} & \textbf{1599.7} & \textbf{1017.3} & \textbf{14.3} & \textbf{8.9} & \textbf{4.9}

\enddata
\tablecomments{
Table columns show the RMSE when comparing the imputation predictions to the true values of the \textit{normalized} features -- \texttt{BWIN\_WORLD}/seeing and \texttt{FLUX\_MAX}/\texttt{FLUX\_APER}$_{n}$, where $n$ is the aperture size in pixels.
 }
\end{deluxetable*}

To test which of these two methods works best for the features with missing values, we perform 3-fold cross validation (CV) run on the training set to estimate the value of the 7 aforementioned features for every source where the feature value is not missing. Thus, we can compare the imputation estimate with the true value and evaluate which of the two methods is superior via the root-mean-square error (RMSE). The results of this test are shown in Table~\ref{tbl:impute}. The performance of the two methods is similar for each of the aperture flux features, however, the RF regression method clearly provides superior predictions for \texttt{BWIN\_WORLD}. We will later show that \texttt{BWIN\_WORLD} is important for star-galaxy classification, thus we adopt the RF regression method for feature imputation.

\subsection{Feature Selection}

In addition to the curation of the training set and choice of algorithm, feature engineering is an important step during machine-learning model construction. As previously noted, RF methods are relatively insensitive to weak or uniformative features, and they also perform well in the presence of strongly correlated features (e.g., \citealt{Richards12a}). Nevertheless, we test the full feature set to see if the model performance can be improved by removing some features. 

RF methods naturally provide a measure of relative feature importance: features that are preferentially selected near the top of individual decision trees contribute more to the final classification predictions than features near the bottom of the trees. Aggregating this information over all trees provides a measure of relative importance for the individual features. In the presence of highly-correlated features, this method does not provide perfectly accurate results as the correlated features may replace each other at the top of the tree thereby suppressing their relative importance. Nevertheless, we employ the RF feature importance rankings to determine which features, if any, can be removed from the model.

\begin{figure}
\centerline{\includegraphics[width=3.3in]{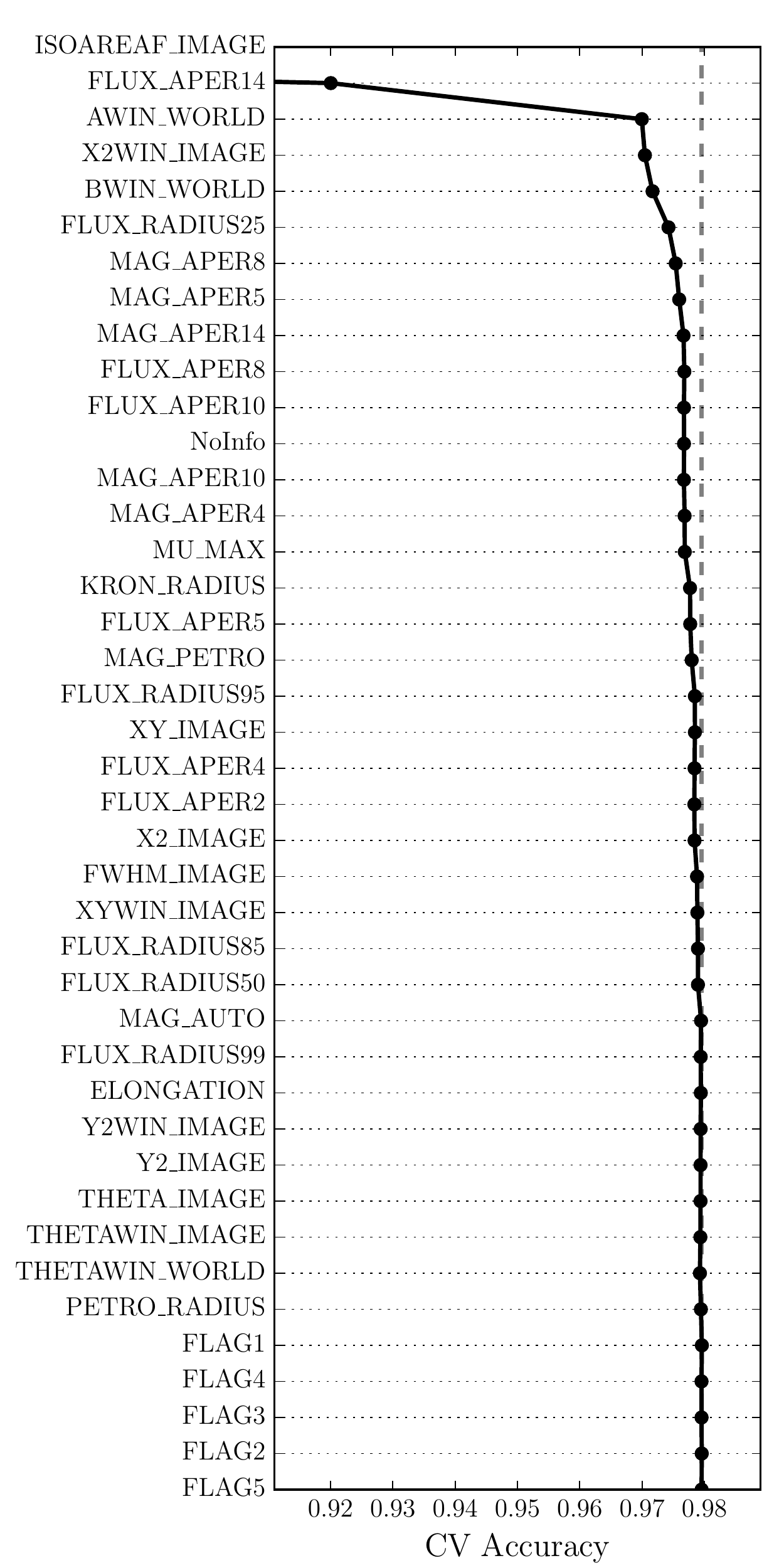}}
\caption[]{Results from the feature selection procedure. Starting from an empty feature set, features are iteratively added in the order of their RF-ranked importance. The model accuracy progressively increases through the 37$^\mathrm{th}$ feature, \texttt{FLAG1}. The features listed below \texttt{FLAG1} are excluded from the final model. The vertical, dashed line shows the optimal CV accuracy.}
\label{fig:feat_import}
\end{figure}

Our procedure is similar to the one employed in \citet{Richards12a}. We construct a series of RF models whereby we iteratively add one feature at a time to each successive model starting with the most-important-RF feature and ending with the 41$^\mathrm{st}$ most important.\footnote{Three of the \texttt{SExtractor} flags were identified as having zero importance by RF, so we excluded them from this exercise. We further added an uniformative feature, \texttt{NoInfo}, which is identically 0 for all sources. The inclusion of \texttt{NoInfo} can help to identify noisy features (see \citealt{Brink13}).} We assess the accuracy of each model via 3-fold CV run on the training set, and we repeat this procedure 5 times to estimate the scatter in the performance of each model. The results of this procedure are shown in Figure~\ref{fig:feat_import}, which shows that only 3 features are needed to achieve a classification accuracy within $\sim$1\% of the maximum CV accuracy. The gains in accuracy beyond the first 3 features are marginal, but increasing, through the 37$^\mathrm{th}$ feature, \texttt{FLAG1}, after which the accuracy decreases slightly. Thus, we select all the ranked features up to \texttt{FLAG1} for use by the final classification model. The exclusion of the last four features does not significantly alter the final model predictions. 

Finally, we note that we explored the procedure described in \citet{Dubath11} to only include uncorrelated features in the final RF model. This procedure produced a maximum CV accuracy of 0.965, whereas the method described above produces a maximum CV accuracy of 0.980. Given the $\sim$1.5\% improvement in prediction accuracy, we elect to include the correlated features in the final model. 

\subsection{Optimizing the Model Tuning Parameters}\label{sec:tuning_params}

The RF algorithm has multiple tuning parameters, which, in combination, control the smoothness of the model projection in the multidimensional feature space. The two most important tuning parameters, $N_\mathrm{tree}$ and $m_\mathrm{try}$, were previously mentioned in \S~\ref{sec:RFalgorithm}. Additional tuning parameters control the depth of individual trees in the forest. We optimize the \texttt{nodesize} parameter, which prevents further splitting of the tree if it would result in a terminal node with fewer than \texttt{nodesize} sources. 

To optimize the model, we perform a coarse-grid search over the three tuning parameters. At each point on the grid, we perform 3-fold CV on the training set to evaluate the model accuracy for the given tuning parameters. We further refine the tuning parameters using a fine-grid search centered on the optimal model from the coarse-grid search. The final optimized model has a CV accuracy of 0.98, with parameters $N_\mathrm{tree} = 750$, $m_\mathrm{try} = 14$, and \texttt{nodesize} = 1. We note that the final model predictions are not sensitive to the tuning parameters: the worst model from the fine-grid seach is $<0.1\%$ worse than the optimal model.

\section{Evaluation of the Optimized Model}\label{sec:results}

\subsection{PTF RF, \texttt{CLASS\_STAR}, and SDSS Comparison }

To test the accuracy of the final, optimized model, we train a RF using the training set and the optimized tuning parameters from \S\ref{sec:tuning_params}. This model is then applied to the independent test set, where we can compare the model predictions to spectroscopic classifications. For the test set, the RF model produces an overall prediction accuracy of 98.0\%, which represents a a $\sim$3.9\% improvement over the accuracy of the \texttt{SExtractor} stellarity measure \texttt{CLASS\_STAR} (94.4\%).\footnote{Overall model accuracies are evaluted using a threshold of 0.5 for separating stars and galaxies. Thus, an RF probability $> 0.5$ or \texttt{CLASS\_STAR} $> 0.5$ results in a stellar classification for the PTF RF model and \texttt{SExtractor}, respectively.} More impressive, however, is the performance of the RF model on faint sources. For test set sources with $r' \ge 21$ mag, \texttt{CLASS\_STAR} has an accuracy of 77.2\% while the RF model has an accuracy of 92.0\%. This represents an improvement of $\sim$19\% for the faintest sources detected by PTF. 

Interestingly, neither method performs as well as the simple parametric method employed by the SDSS pipeline. In brief, the SDSS pipeline identifies all sources with \texttt{psfMag} - \texttt{cModelMag} $> 0.145$ as galaxies \citep{Lupton02}, where \texttt{psfMag} is the point-spread-function magnitude and \texttt{cModelMag} is the composite model magnitude resulting from the best-fitting linear combination of the best-fitting de Vaucouleurs and exponential model for an object's light profile. For the test set, the SDSS photometric classification provides an overall accuracy of 99.6\% and an accuracy of 99.1\% for sources with $r' \ge 21$ mag. We attribute the improved performance of the simplistic SDSS photometric classifier to their higher quality observations, including better seeing $\sim$1.4$\arcsec$ for SDSS \citep{Abazajian03} vs.\ $\sim$2.4$\arcsec$ for PTF and greater depth for SDSS.\footnote{The SDSS photometric classifier uses the sum of \texttt{psfMag} - \texttt{cModelMag} across all 5 filters to perform the final star-galaxy classification.}

\begin{figure}
\centerline{\includegraphics[width=3.5in]{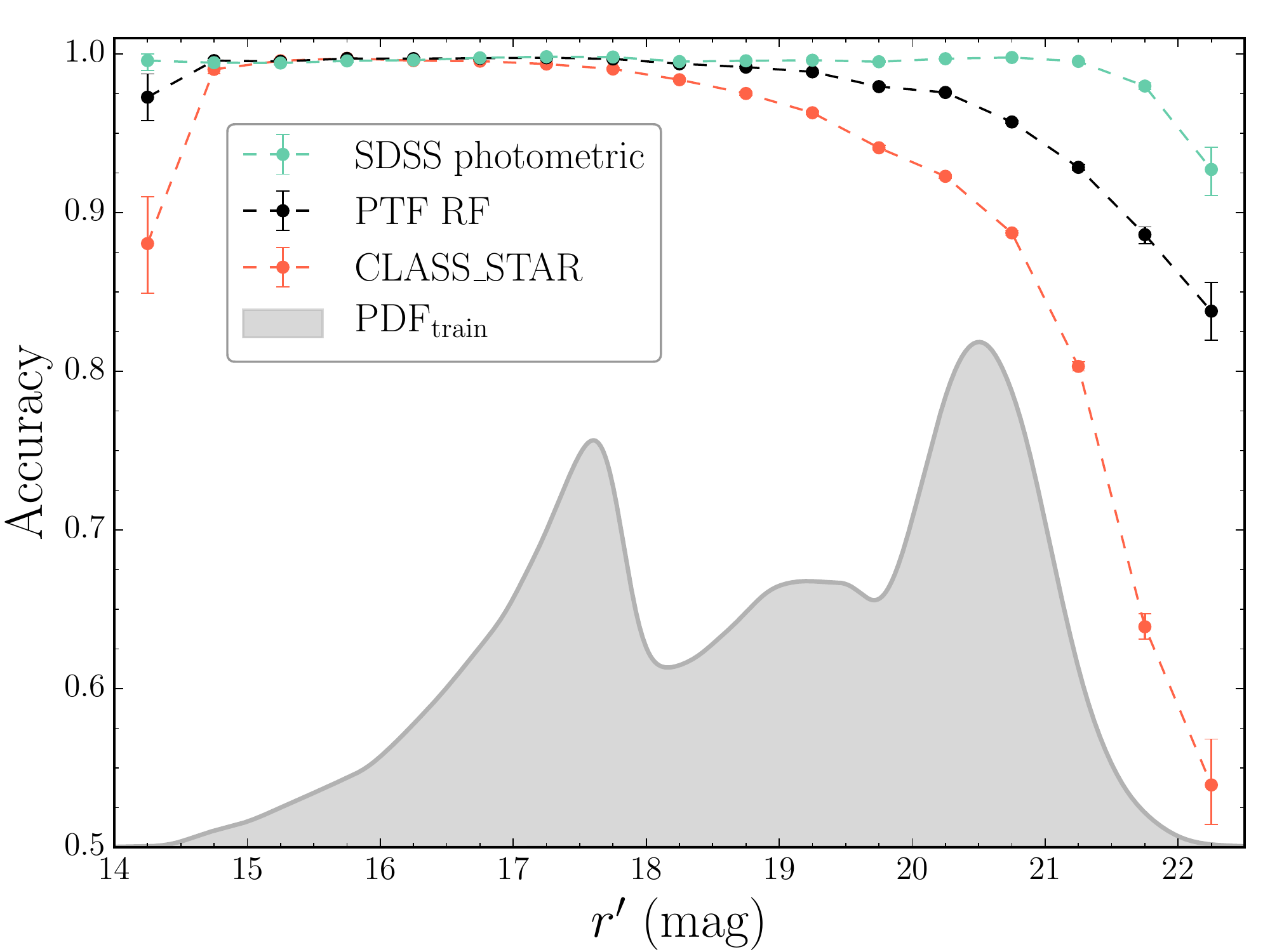}}
\caption[]{Photometric classification accuracy for \texttt{SExtractor}/\texttt{CLASS\_STAR} (pink points), the PTF RF model (black points), and the SDSS photometric pipeline (light green points) as a function of magnitude for all sources in the test set. A KDE of the PDF of $N(m)$ for training set sources is shown in grey. Accuracies are shown in bins of width 0.5 mag, and the error bars reflect the 95\% confidence interval on the mean accuracy from 500 bootstrap resamples in each bin. The PTF RF model shows a significant improvement over \texttt{CLASS\_STAR}, especially at faint magnitudes.}
\label{fig:sdss_ptf_se_comp}
\end{figure}

In Figure~\ref{fig:sdss_ptf_se_comp}, we compare the accuracy of the RF model, \texttt{CLASS\_STAR}, and the SDSS photometric classifier evaluated via the test set as a function of magnitude. Figure~\ref{fig:sdss_ptf_se_comp} shows that the performance of each method is similar down to $r' \approx 18$ mag. This is to be expected based on Figure~\ref{fig:sG_rogues}, which shows that galaxies and stars are clearly separable over this magnitude range in PTF imaging. The performance of \texttt{CLASS\_STAR} quickly degrades for fainter sources, however, to the level that \texttt{CLASS\_STAR} is similar to random guessing for sources with $r' > 22$ mag. As previously noted, the RF model provides superior predictions for faint sources, which will enable us to better identify stars in PTF imaging outside the SDSS footprint. Figure~\ref{fig:sdss_ptf_se_comp} also shows a KDE of the PDF of $N(m)$ for training set sources, which is virtually identical to the PDF for test set sources. The RF model is most reliable in regions of high density, roughly $16.5 \, \mathrm{mag} \lesssim r' \lesssim 21.3 \, \mathrm{mag}$.

\subsection{RF Model Accuracy for Stars and Galaxies}\label{sec:starGalAcc}

\begin{figure*}
\centerline{\includegraphics[width=6.5in]{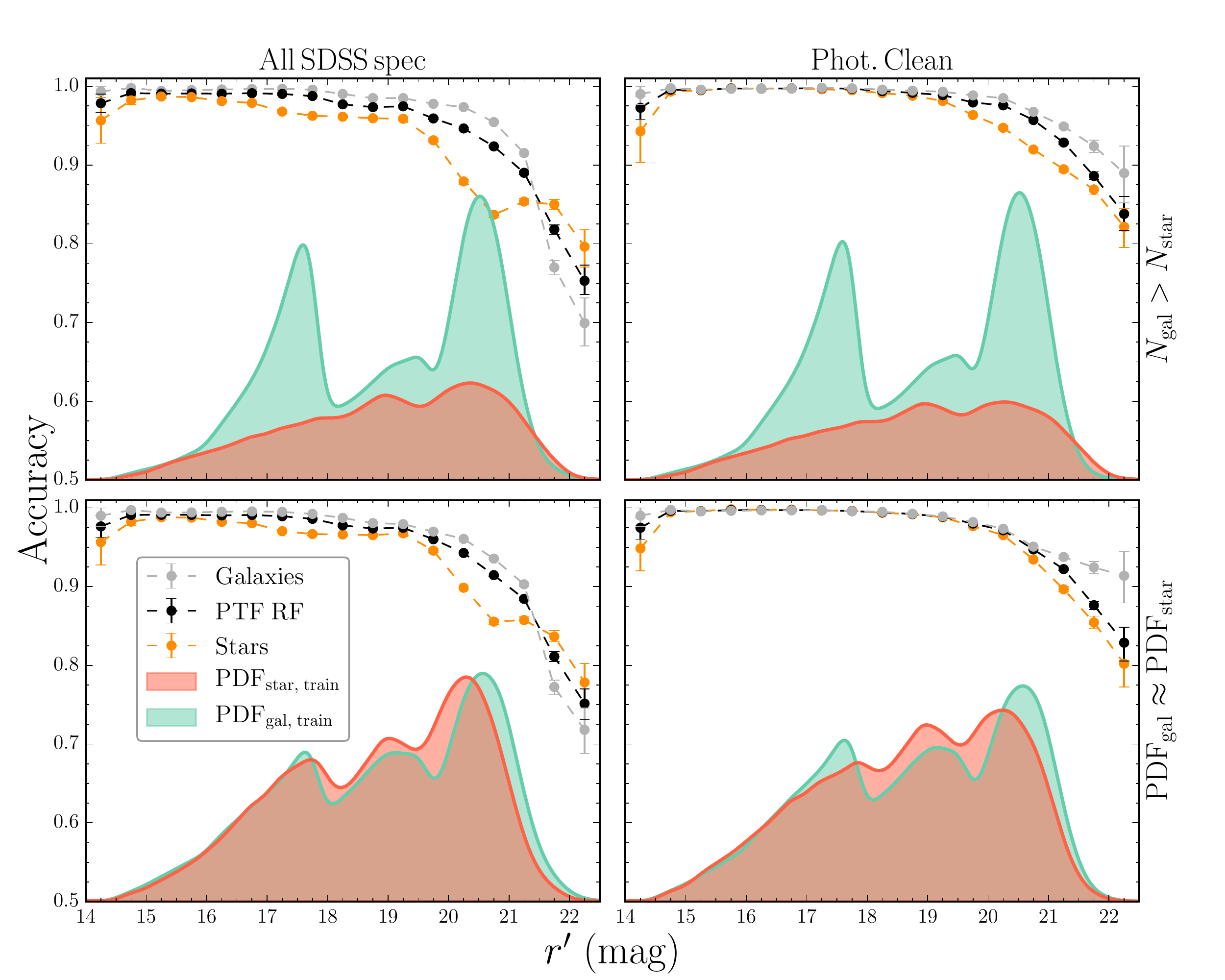}}
\caption[]{Accuracy of the PTF RF test set predictions as a function of magnitude for different permutations of the training set. In each panel, the black points show the overall accuracy of the model, the grey points show the accuracy when only considering galaxies, and the orange points show the accuracy for stars. Additionally, a KDE of the PDF for stars and galaxies in the training set are shown in pink and light green, respectively. The stellar PDF has been normalized by the ratio of the number of stars to the number of galaxies in the training set. The training set variations are as follows: \textit{upper left} -- full training set including all PTF sources with SDSS spectra, \textit{lower left} -- balanced version of the full training set designed to have $\mathrm{PDF}_\mathrm{gal} \approx \mathrm{PDF}_\mathrm{star}$ (see text for further details), \textit{upper right} -- the \textit{photometrically clean} training set (see \S\ref{sec:systematics}), and \textit{lower right} -- balanced \textit{photometrically clean} training set. See Figure~\ref{fig:sdss_ptf_se_comp} for a definition of the bin width and uncertainties.
}
\label{fig:stars_galaxies}
\end{figure*}

As previously noted, the primary motivation for constructing a PTF star-galaxy catalog is to identify a pristine list of point sources in PTF imaging. We are particularly interested in the accuracy with which we can identify faint stars as these are the most likely false positives in the search for fast-transients \citep{Kulkarni06, Berger13}. Similar to above, we plot the accuracy of the RF model for classifying stars and galaxies in the test set as a function of brightness in Figure~\ref{fig:stars_galaxies}. 

Figure~\ref{fig:stars_galaxies} features 4 panels, each of which reflects slight variations on the training set. The left column shows training sets that include all PTF sources with SDSS spectra, while the right column shows results from the \textit{photometrically clean} training set (see \S\ref{sec:systematics}). The top row shows training sets including all available stars and galaxies, while the bottom row shows the results when the stars and galaxies in the training set are downsampled such that both classes have similar PDFs of magnitude. 

The upper-left panel contains the most striking feature in Figure~\ref{fig:stars_galaxies}: the kink in the accuracy curve for stars at $r' \approx 20.7$ mag, followed by the crossing of the star and galaxy accuracy curves at $r' \approx 21.5$ mag. A less significant, but nonetheless noticable kink also appears near $r' \approx 17.5$ mag in the stars accuracy curve. These departures from a smooth accuracy curve occur near peaks in the galaxy PDF, where stars are most significantly outnumbered.  

Initially, we believed the kinks could be removed by balancing the magnitude PDF for stars and galaxies in the training set. To achieve this balance, we use KDEs of the PDFs from $r' = 13.5$ mag to $r' = 22.5$ mag. The stellar KDE and galaxy KDE are evaluated at the brightness of each galaxy, and the former is divided by the latter to provide a weight. We then select a weighted random sample of 500,000 galaxies for the balanced training set. We use the same procedure to select a weighted random sample of 500,000 stars, however, the weights are determined by dividing the galaxy PDF by the stellar PDF. All training set sources with brightness outside the range $r' = 13.5-22.5$ mag are also included, resulting in a final balanced training set with 1,001,975 sources. The balanced training set PDFs and accuracy curves are shown in the lower-left panel of Figure~\ref{fig:stars_galaxies}. While the significance of the kinks is reduced when using the balanced training set, it is clear that balancing the two classes does not eradicate this unusual systematic behavior.

Ultimately, the kink in the upper-left panel is due to the SDSS targeting bias toward LRGs, a small fraction of which turn out to be photometrically-blended red stars, as previously noted in \S\ref{sec:systematics}. In brief, the removal of stars targeted as galaxies, galaxies targeted as stars, and low-$z$ QSOs from the training and test sets dramatically improves the performance of the star-galaxy classification model.\footnote{For the full details on which sources are removed from the training and test set, see \S\ref{sec:systematics}.} This also removes the unusual systematics from the accuracy curves, as seen in the right column of Figure~\ref{fig:stars_galaxies}. We applied the same procedure described above to balance the \textit{photometrically clean} training set, and the resulting predictions are shown in the lower-right panel of Figure~\ref{fig:stars_galaxies}. Ultimately, the performance of the full and balanced \textit{photometrically clean} training sets was nearly identical with an overall accuracy of 98.0\% and 97.8\%, respectively. Ultimately, we adopt the full \textit{photometrically clean} training set for the final RF model as this provides the most information to the classifier. The use of the balanced \textit{photometrically clean} training set would not significantly alter the final model classifications. 

\subsection{Selecting a Pristine Sample of Stars}

Rather than producing the best overall accuracy possible, we hope to generate a catalog of PTF point sources that is virtually free of galaxies. In addition to providing classifications, RF models also produce relative rankings of the class likelihood for newly observed sources by recording the fraction of trees in which each source is assigned to each class.\footnote{These relative rankings are often referred to as RF probabilities. However, the RF probability score does not represent a true probability as it is a strong function of the training set, which in virtually all astronomical applications is biased relative to the true distributions present in nature. Thus, we prefer to refer to this quantity as the RF relative ranking.} For our two class problem, a source that is classified as a star in every tree would have a RF relative ranking equal to 1, while a source labeled a galaxy in every tree would have ranking 0. Sources with RF relative ranking $\approx$ 0.5 are somewhat ambiguous, with the trees nearly divided on the classification. 

Thresholds can be placed on the RF relative ranking to adjust the overall number of false positives (FP) and true positives (TP) produced by the classifier. Above, a threshold of 0.5 was adopted to test the overall accuracy of the classifier. Now, we adjust that threshold to reduce the number of FP, which for our purposes are considered far more harmful than false negatives (FN). Threshold adjustments are typically determined using a receiver-operating-characteristic (ROC) curve, which plots the true positive rate (TPR),
$$\mathrm{TPR} = \frac{\mathrm{TP}}{\mathrm{TP} + \mathrm{FN}},$$
against the false positive rate (FPR) 
$$\mathrm{FPR} = \frac{\mathrm{FP}}{\mathrm{FP} + \mathrm{TN}},$$
where TN is the number of true negatives, as the classification threshold is varied from 1 to 0. The performance of different models can be compared via ROC curves by examining which comes the closest to the classification ideal of $\mathrm{TPR} = 1$ and $\mathrm{FPR} = 0$.

\begin{figure}
\centerline{\includegraphics[width=3.5in]{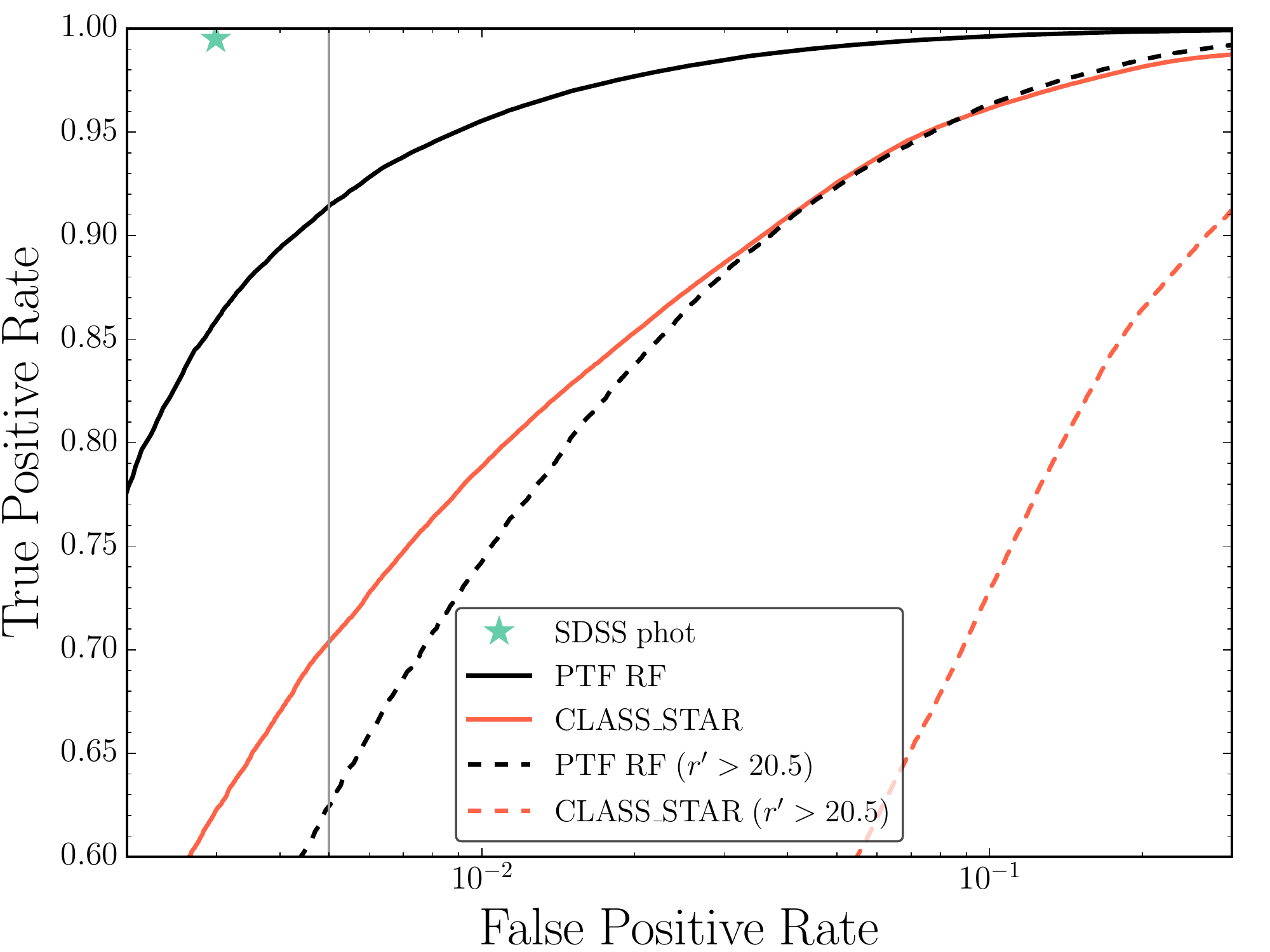}}
\caption[]{ROC curves comparing the relative performance of SDSS, the PTF RF model, and \texttt{SExtractor}. The solid black and red lines show the ROC curve for the PTF RF model and \texttt{SExtractor}, respectively, as evaluated by the \textit{photometrically clean} test set. The \texttt{SExtractor} ROC curve is generated by varying the classification threshold from \texttt{CLASS\_STAR} = 0 to 1. The dashed black and red lines show the ROC curves for faint ($r' > 20.5$ mag) \textit{photometrically clean} test set sources for PTF and \texttt{SExtractor}, respectively. The solid vertical line shows the desired FPR = 0.005 for the final PTF point-source catalog. The SDSS classifier is shown as a turquoise star due to the binary nature of the SDSS photometric classification.
}
\label{fig:ROC}
\end{figure}

ROC curves for the PTF RF model and \texttt{SExtractor} are shown in Figure~\ref{fig:ROC}. For \texttt{SExtractor} the curve is determined by varying the classification threshold from \texttt{CLASS\_STAR} = 0 to 1. The performance of SDSS is shown as a single point because the SDSS pipeline provides a single binary classification without any information on the relative likelihood for individual sources. Similar to Figure~\ref{fig:sdss_ptf_se_comp}, Figure~\ref{fig:ROC} shows that the PTF RF model significantly outperforms the \texttt{SExtractor} model. In fact, the performance of the PTF RF model on faint sources ($r' > 20.5$ mag) is virtually identical to the performance of \texttt{SExtractor} on the entire test set. As has already been noted, the superior quality of SDSS imaging results in higher fidelity classifications than is possible with PTF imaging. 

The classification threshold adopted for the PTF point-source catalog is determined by maximizing the TPR at a $\mathrm{FPR} = 0.005$. The adoption of this low FPR ensures that less than 0.5\% of galaxies will be included in the point-source catalog and thereby excluded from examination should they host a transient. For the test set at $\mathrm{FPR} = 0.005$, the PTF RF model produces a $\mathrm{TPR} = 0.91$, corresponding to a classification threshold of 0.83 for the RF relative ranking. Below, we show that the performance of the model as measured by the test set likely overstates the model accuracy when applied to sources in the field. Thus, we ultimately adpot a classification threshold that is more conservative than 0.83. 

\section{Implementing the Catalog}

\subsection{Final Field-source Predictions}\label{sec:field_preds}

The final step for incorporating the star-galaxy catalog into the appropriate PTF pipelines is to apply the RF model to the 548,687,903 sources detected in PTF reference images. To assess the efficacy of the model as applied to the field star population, which is dominated by sources at the faint end of the test set, we compare our final PTF classifications to those made by the SDSS photometric pipeline, which for this purpose we adopt as ground truth. 

\begin{figure*}
\centerline{\includegraphics[width=7in]{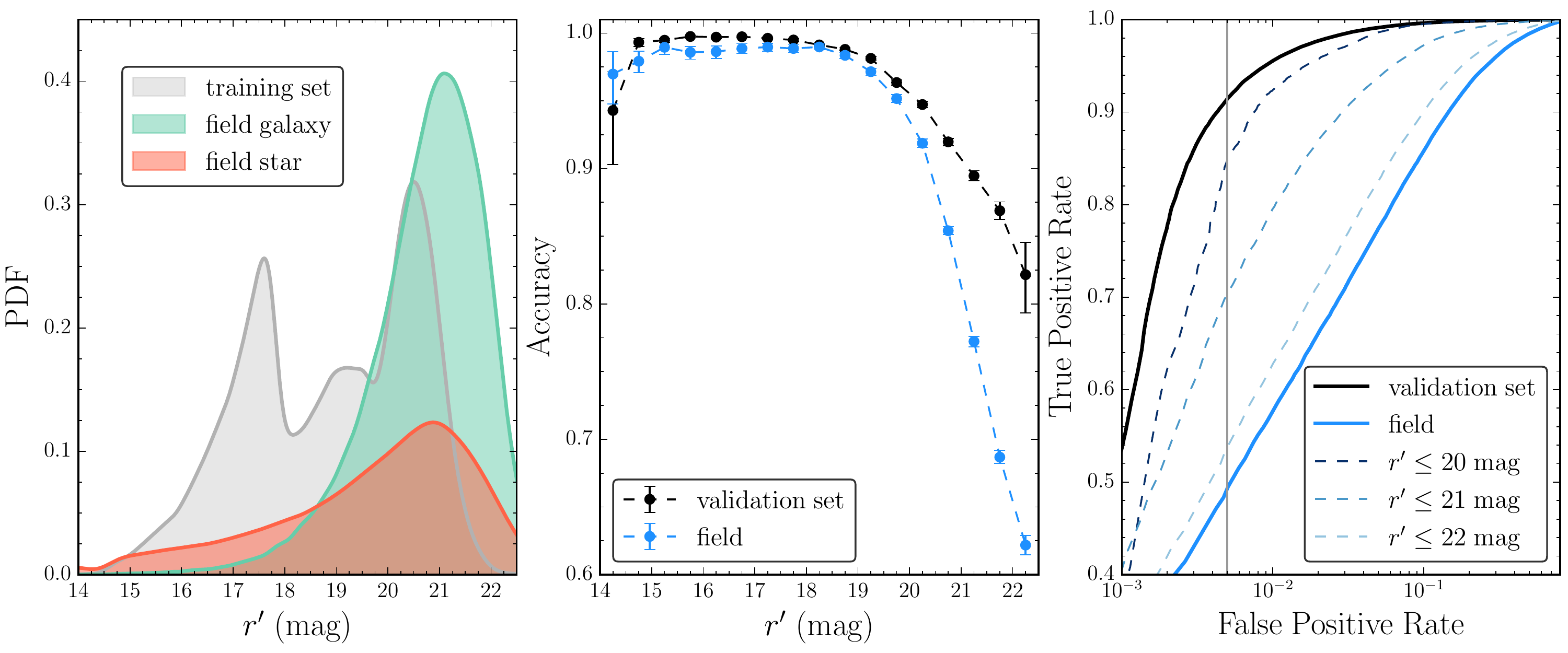}}
\caption[]{Accuracy and ROC curves for the PTF RF model compared to SDSS photometric classifications. \textit{Left} -- KDEs of the magnitude PDF for the full training set (grey), SDSS photometric stars (pink), and SDSS photometric galaxies (light green). \textit{Middle} -- accuracy curves for the PTF RF model as tested on the test set (black points) and tested by the SDSS photometric classification (light blue points). The performance on the random set of field sources shows that the test set predictions overstate the true accuracy of the model. \textit{Right} -- ROC curve for the test set predictions (solid-black line) and the predictions for SDSS field sources (solid-blue line). The dashed lines show ROC curves when constraining the field sample to stars brighter than 20, 21, and 22 mag. The solid vertical line shows the FPR = 0.005 cut adopted for inclusion in the final PTF point-source catalog.
}
\label{fig:field}
\end{figure*}

To test the performance of the model on the field, we randomly select 300,000 sources with $\mathrm{RA}_{\mathrm{J}2000}$ between $12^{\mathrm{h}}$ and $13^{\mathrm{h}}$ and $\mathrm{Dec}_{\mathrm{J}2000}$ between $20^{\circ}$ and $35^{\circ}$ from the PTF reference image catalogs. This area was selected to test the model at high galactic latitudes ($b > 75^{\circ}$); we expect blending to be significantly worse near the galactic plane ($\left|b\right| \lesssim 15^{\circ}$), which will in turn degrade the quality of the model. Using SDSS \texttt{CasJobs} we perform a 1$\arcsec$ crossmatch between the randomly selected sources and the SDSS photometric catalog, yielding 280,972 common sources, which we refer to as the SDSS field set. Our RF model predictions produce an overall accuracy of 83.8\% when compared to the SDSS photometric classifications, $\sim$15\% worse than the $\sim$98\% accuracy reported for the test set (\S\ref{sec:starGalAcc}). This degredation in performance is expected as the field population is much fainter than the test set. 

Figure~\ref{fig:field} shows accuracy and ROC curves to compare the performance of the model on the test set versus the field. The left panel of the figure illustrates that the typical field source is significantly fainter than those present in the training/test set. The middle panel shows that predictions on the test set overstate the accuracy of the model for sources with $r' \gtrsim 20$ mag. This is further corroborated by the right panel, which shows that the test set ROC curve and the ROC curve for bright field sources ($r' \le 20$ mag) are nearly identical. The ROC curves show successively worse performance when including fainter and fainter sources. 

The one caveat to these conclusions is that the SDSS photometric classifications do not truly provide ground truth: Figure~\ref{fig:sdss_ptf_se_comp} shows that the accuracy of the SDSS model drops to $\sim$93\% near the PTF reference image detection limit. Furthermore, the \textit{photometrically clean} training and test sets overstate the accuracy of all models as photometric blends have been actively removed. Nevertheless, the results presented in this section are comparative. Sources that are blended in SDSS imaging should also be blended in PTF imaging, meaning that in many of these cases both classifiers will have the same incorrect classification. Thus, the divergence between the two curves in the middle panel of Figure~\ref{fig:field} cannot be explained completely by misclassifications by SDSS at the faint end. 

The final RF relative ranking used to select a pristine sample of point sources is determined from the ROC curves shown in Figure~\ref{fig:field}. Prior to selecting a RF relative ranking corresponding to $\mathrm{FPR} = 0.005$, we impose a magnitude cut, $R_\mathrm{PTF} \le 21 \; \mathrm{mag}$, as the accuracy of the SDSS photometric classifier quickly declines for $r' > 21 \; \mathrm{mag}$.\footnote{See Figure~\ref{fig:sdss_ptf_se_comp} and \url{http://www.sdss.org/dr12/imaging/other_info/\#stargalaxy}.} There are 463,581,596 PTF sources with $R_\mathrm{PTF} \le 21 \; \mathrm{mag}$, and the $\mathrm{TPR} = 0.695$ at $\mathrm{FPR} = 0.005$ for this subset of the data. This corresponds to an RF relative ranking threshold of 0.966, meaning only sources classified as point sources in $\ge$725 trees in the forest pass the cut. While as many as $\sim$30\% of the \textit{true} $R_\mathrm{PTF} \le 21 \; \mathrm{mag}$ point sources are missed by this cut, our objective is to create a catalog of point sources with virtually no galaxies classified as stars. The vast majority of sources are faint, where classification is the most challenging, meaning this requirement results in a final point-source catalog that is incomplete. Application of the 0.966 threshold yields a final point-source catalog containing 170,440,636 $R_\mathrm{PTF} \le 21 \; \mathrm{mag}$ sources, $\sim$30\% of \textit{all} sources extracted from the PTF reference images. 

\subsection{Comparison to the Previous Star Catalog}\label{sec:nersc_cat}

Prior to the completion of the RF point-source catalog, the PTF real-time pipeline \citep{Cao16} utilized a star catalog based on several cuts on \texttt{SExtractor} parameters. Hereafter, we refer to this initial star catalog as the NERSC catalog. Real-time transient candidates that are spatially coincident with sources in the NERSC catalog are rejected as false positives and removed from the stream prior to human vetting. All NERSC reference-image sources\footnote{The reference images utilized in this study are, on average, slightly deeper than NERSC pipeline references. The same procedure is used to create both sets of references, after which \texttt{SExtractor} is used for source detection.} satisfying the following cuts:
\begin{flalign*}
    &\texttt{MAG\_BEST} < 20,&\\ 
    &\texttt{MU\_MAX} - \texttt{MAG\_BEST} < 0.2 + \mathrm{med}(\texttt{MU\_MAX} - \texttt{MAG\_BEST}),& \\
    &\texttt{FWHM\_IMAGE} < 2 \times \mathrm{med}(\texttt{FWHM\_IMAGE}),&\\
    &\texttt{ELONGATION} < 1.5,&
\end{flalign*}
where $\mathrm{med}()$ refers to the median value for all sources detected on the same CCD, are classified as stars in the NERSC catalog. Initial testing showed that these cuts identified point-sources more reliably than \texttt{CLASS\_STAR}. 

To compare the performance of the NERSC catalog and the PTF RF catalog, we adopt the SDSS field set from \S\ref{sec:field_preds}. Using a 1$\arcsec$ radial crossmatch, there are 241,675 sources in common between the SDSS field set, the NERSC catalog, and the PTF RF catalog. As the NERSC catalog adopts a single hard cut for classification, a comparison of ROC curves is not possible. Instead, we compare the confusion matrix for each, which summarizes the total number of stars classified as stars (TP), galaxies classified as stars (FP), galaxies classified as galaxies (TN), and stars classified as galaxies (FN). Ideally, the confusion matrix would only have power along the diagonal, indicating perfect classification. The PTF RF catalog, however, has been optimized to minimize FP (by adopting a classification threshold $>$ 0.966), resulting in significant off-diagonal power. 

\begin{figure}
\centerline{\includegraphics[width=3.5in]{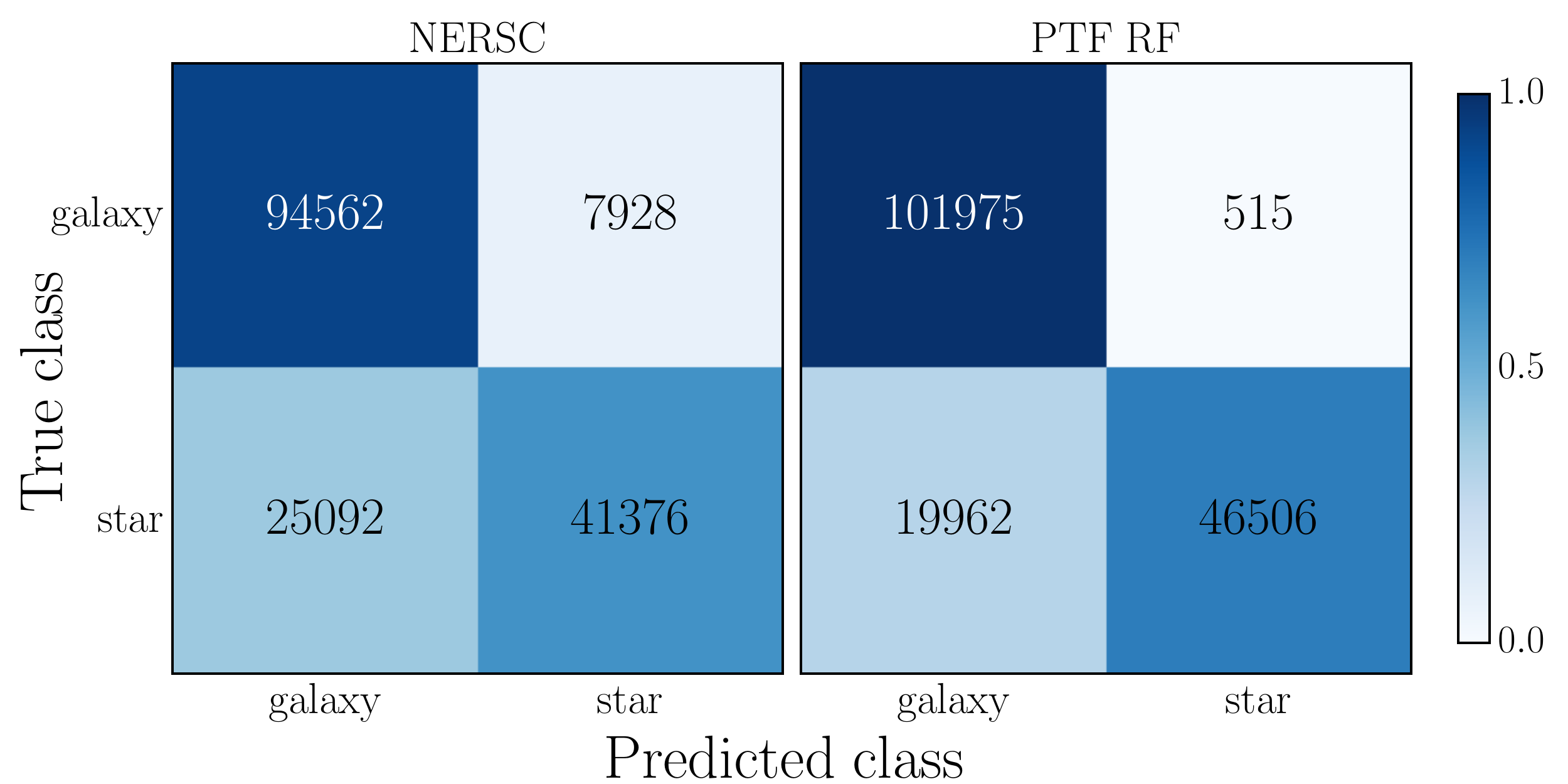}}
\caption[]{Confusion matrix comparison between the NERSC catalog and the PTF RF catalog. Each matrix shows (clockwise, from the upper left) the total TN, FP, TP, and FN. The colors represent the fraction of true class members. The PTF RF catalog has a $\sim$12\% improvement in TPR, and, more importantly, a factor $\sim$15 decrease in the FPR relative to the NERSC catalog.
}
\label{fig:NERSCvIPAC}
\end{figure}

Limiting the sample to sources with $R_\mathrm{PTF} \le 21 \; \mathrm{mag}$ and adopting the SDSS photometric classifications as ground truth yields the confusion matrices shown in Figure~\ref{fig:NERSCvIPAC}. The shading in each matrix shows that the qualitative performance of the catalogs is similar. In detail, however, the PTF RF catalog produces more TP, and most importantly, a factor of $\sim$15 fewer FP. The NERSC catalog removes $\gtrsim$7.5\% of all galaxies from the search for transients. The PTF RF catalog reduces this fraction to $\lesssim$0.5\%, while also rejecting a larger number of true point-sources from the candidate stream. Thus, adoption of the new PTF RF catalog significantly improves the search for all transients relative to the NERSC catalog.

\subsection{Demonstrable Improvements in the Discovery Potential of the PTF RF Catalog}

While \S\ref{sec:nersc_cat} provides evidence that $\sim$7\% of galaxies are misclassified as stars by the NERSC catalog, here we provide definitive examples of transients PTF missed that would have been detected had the PTF RF catalog been employed. These transients were identified via a \textit{non-exhaustive} search, which included the following steps:

\begin{enumerate}
    
    \item All transient candidates with the NERSC flag \texttt{is\_star} AND satisfying the normal thresholds for human vetting during the period from 2015 Nov 01.0 UT to 2016 Jan 01.0 UT were selected. This selection yielded 72,546 unique sources that were detected between 2 and 170 times during the search period.
    \item These $\sim$72k sources were cross-matched against the PTF RF catalog to identify candidates classified as stars in the NERSC catalog and galaxies in the PTF RF catalog, resulting in a list of 25,138 sources.
    \item Those candidates with detections before 2015 Aug 01.0 UT or after 2016 Mar 01.0 UT were removed to exclude long-term variables, which reduced the list to 15,737 candidates.
    \item These were cross-matched to SDSS to provide color and morphological information, further culling the list to $7{,}813$ sources.
    \item The 39 sources classified as galaxies by SDSS with $\ge 5$ detections between 2015 Nov 01.0 UT and 2016 Jan 01.0 UT and $u_0' - g_0' > 0.7~\mathrm{mag}$ were visually inspected. The color cut was applied to eliminate likely QSOs (see e.g., Fig~4 in \citealt{sesar07}).
\end{enumerate} 

\begin{figure}
\centerline{\includegraphics[width=3.5in]{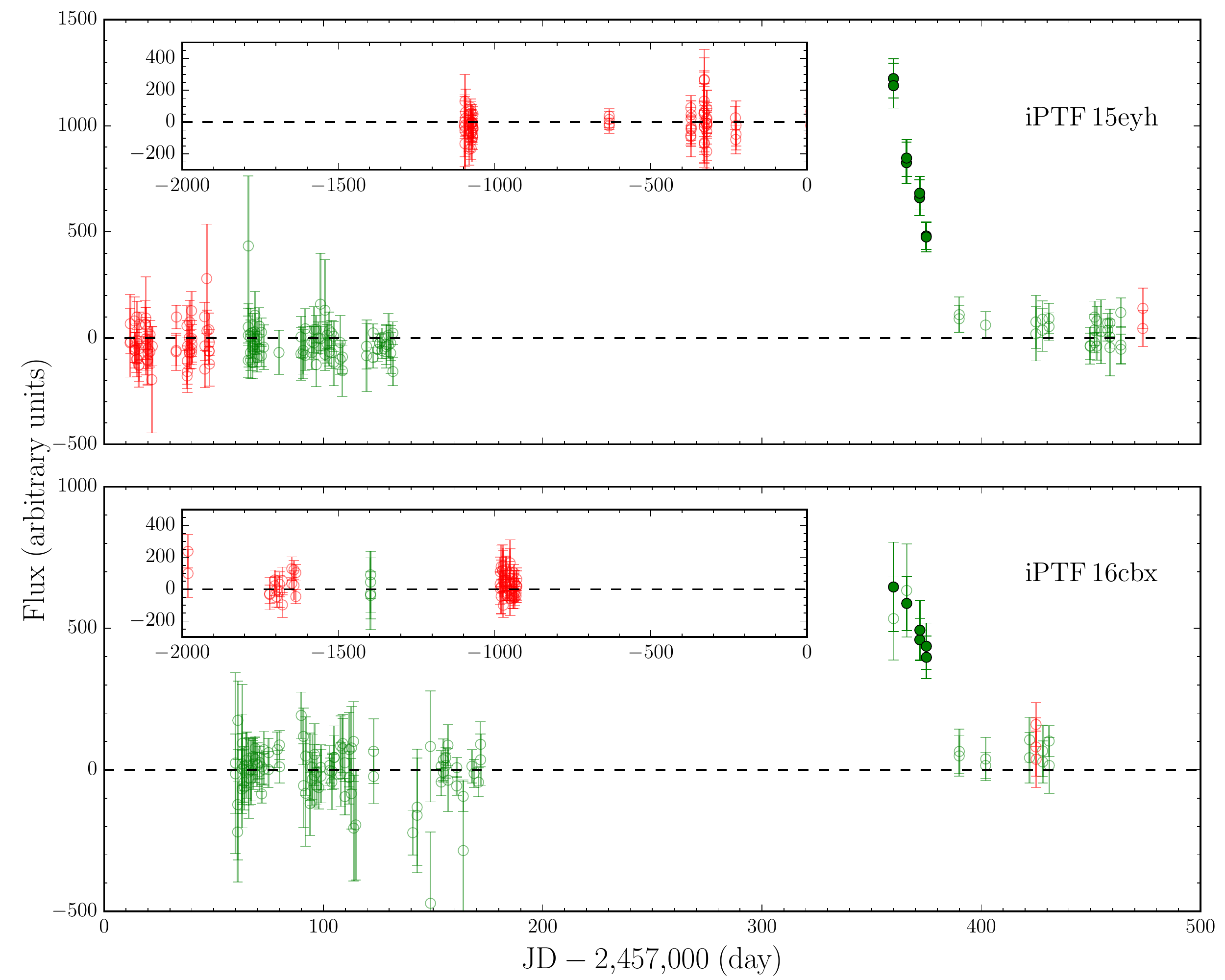}}
\caption[]{
Light curves for the 2 transients missed by the NERSC pipeline, produced via difference-image-PSF photometry at the location of the transient. PSF flux measurements are shown with arbitraty units. $g$-band observations are shown in green, while $R$-band observations are shown in red. Epochs where the transient is detected, i.e. signal-to-noise ratio $\ge 4$, are shown with solid, filled circles, while epochs with no detection are shown with light, empty circles. Both transients were detected over a $\sim$2 week period starting on $\sim$2015 Dec 01 UT. The inset panels show the lack of historical variability over the duration of PTF observations. 
}
\label{fig:tranLCs}
\end{figure}

Visual inspection of these sources revealed 2 transient candidates that were otherwise missed by the NERSC discovery pipeline. These candidates have been internally designated as iPTF~15eyh\footnote{PTF transients are named based on the year when they are discovered, i.e.\ when a human manually saves a candidate as real. In late 2015, the PTF IPAC pipeline (see \citealt{Masci16}) used a preliminary version of the PTF RF catalog to reject stars, and as a result iPTF~15eyh was successfully identified in real-time. We include it here because the NERSC pipeline missed this transient.} and iPTF~16cbx, and their light curves are shown in Figure~\ref{fig:tranLCs}. The lack of historical variability and host galaxy colors suggest these candidates are bonafide transients and not active galactic nuclei. The nature of these transients is difficult to discern given the partial light curve coverage and lack of spectroscopic observations. Nevertheless, our limited, non-exhaustive search reveals that PTF missed several transients due to misclassifications in the NERSC catalog. The use of the PTF RF catalog will significantly lower the number of transients that are missed because their host galaxies are classified as stars.

\subsection{Supplementing the Catalog with SDSS}

It is possible to further improve the rejection of point-sources from the candidate transient stream using SDSS imaging data, which is limited to $\sim$half the total PTF imaging footprint. As previously discussed, SDSS has superior imaging quality to PTF and provides superior photometric classifications (see Figure~\ref{fig:sdss_ptf_se_comp}). The unfiltered addition of SDSS stars to the PTF point-source catalog will reduce its effectiveness, however, as the SDSS classification has not been tuned to produce an $\mathrm{FPR} = 0.005$. Below, we adjust the SDSS classification threshold to produce the desired FPR.  

The SDSS pipeline classifies a source as a star if 
$$f_\mathrm{PSF}/f_\mathrm{cmodel} > 0.875,$$ 
where $f_\mathrm{PSF}$ is the PSF flux, and $f_\mathrm{cmodel}$ is the composite-model flux, which measures the best-fit linear combination of an exponential and a de Vaucouleurs profile. The final classification is performed using the sum of the fluxes in all bands where the source is detected. Adjusting the decision threshold up or down decreases or increases FP, respectively.

\begin{figure}
\centerline{\includegraphics[width=3.5in]{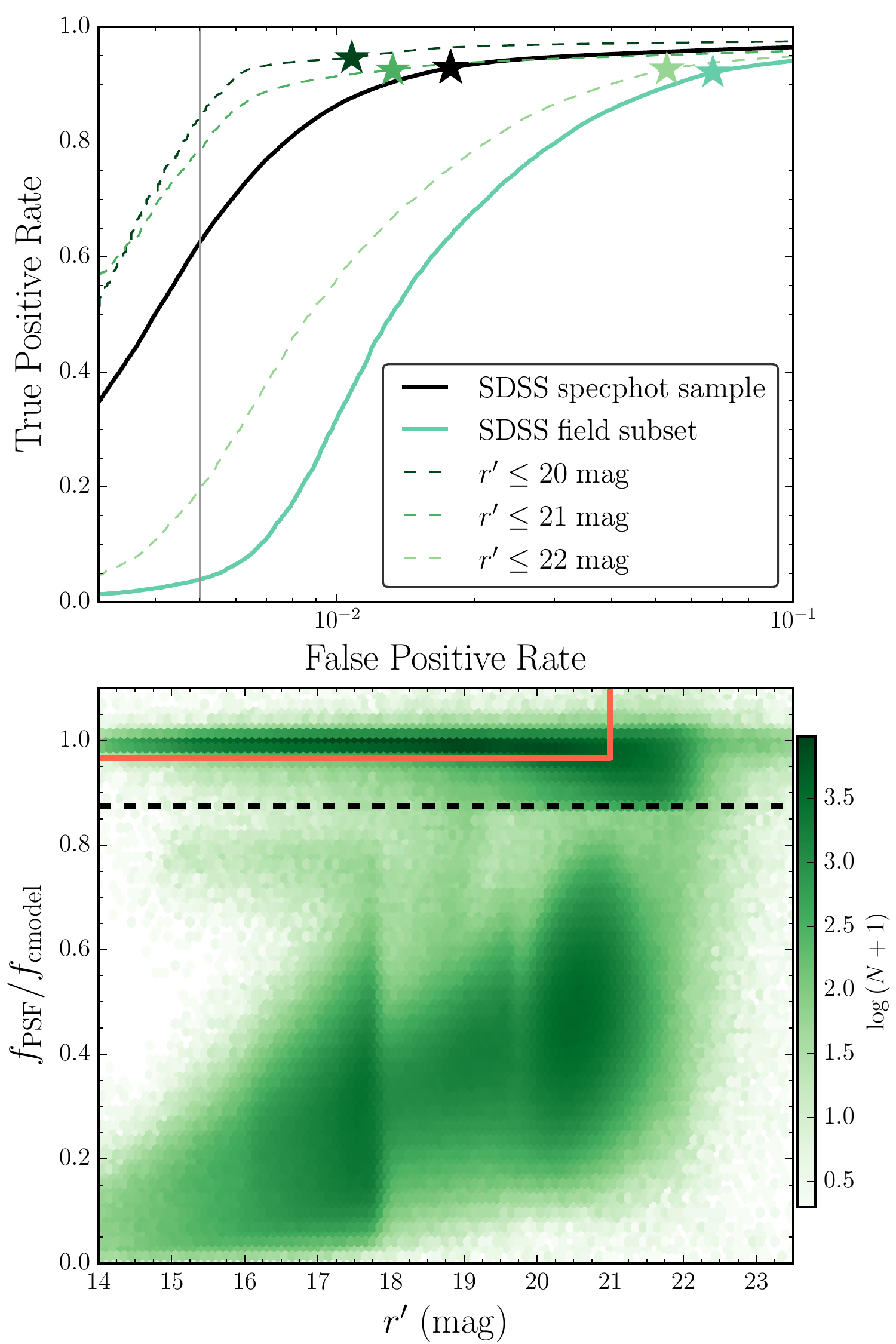}}
\caption[]{Decision thresholds for selecting SDSS point sources with $\mathrm{FPR} = 0.005$. \textit{Top}: ROC-like curves (see text) for SDSS photometric classification. For each curve the solid star marks the location corresponding to $f_\mathrm{PSF}/f_\mathrm{cmodel} = 0.875$, the SDSS pipeline classification threshold. The solid black and light green lines show the SDSS specphot sample and field subset, respectively. The dashed lines show the ROC-like curves for the SDSS field subset restricted to sources brighter than $r' = 20$, 21, and 22 mag from top to bottom. For our purposes, the 0.875 threshold produces too many misclassified galaxies. \textit{Bottom}: Density plot showing $f_\mathrm{PSF}/f_\mathrm{cmodel}$ as a function of $r'$ for all sources in the SDSS specphot sample. Pixels are $\sim$0.1 mag wide. The concentration at $f_\mathrm{PSF}/f_\mathrm{cmodel} \approx 1$ corresponds to point sources. The dashed, horizontal line represents the SDSS classification threshold. Only sources contained by the solid pink lines are selected to supplement the PTF RF point-source catalog. Notice that source classes begin to blend together for $r' > 21$ mag. 
}
\label{fig:SDSS_cuts}
\end{figure}

To determine the optimal threshold for $f_\mathrm{PSF}/f_\mathrm{cmodel}$ we select spectroscopic classifications and photometric fluxes for all SDSS sources. The query is performed via \texttt{CasJobs} to select sources with \texttt{sciencePrimary} = 1 and \texttt{mode} = 1, corresponding to the primary spectroscopic and photometric detection of a given source. A total of 3,537,411 sources match this criteria, which we hereafter refer to as the SDSS specphot sample. We perform an ROC-like analysis to measure changes in the TPR and FPR as a function classification threshold, where we have adjusted $f_\mathrm{PSF}/f_\mathrm{cmodel}$ from its highest value to its lowest. The results of this procedure are shown via the solid black line in Figure~\ref{fig:SDSS_cuts}. The vertical grey line shows the desired $\mathrm{FPR} = 0.005$, while the solid black star shows the location on the curve corresponding to $f_\mathrm{PSF}/f_\mathrm{cmodel} = 0.875$. Thus, adopting the SDSS classification threshold would yield a $\mathrm{TPR \approx 0.93}$ and $\mathrm{FPR} \approx 0.02$, which is significantly higher than our target. It is clear that an alternative threshold is needed to achieve a $\mathrm{FPR} = 0.005$.

The SDSS specphot sample is heavily biased by the SDSS spectroscopic-targeting function, and as such does not reflect the true distribution of SDSS photometric detections. This is illustrated by the grey distribution in the left panel of Figure~\ref{fig:field},\footnote{Strictly speaking Figure~\ref{fig:field} shows the distribution of spectroscopic training sources for the PTF RF model, which is virtually indistinguishable from the SDSS specphot sample.} as compared to the pink and light green distributions. The optimal SDSS classification threshold should be selected from a set of sources that reflect the true distributions found in nature. We approximate such a set of sources via a weighted random subset of the SDSS specphot sample. The individual weights are determined via KDEs of the magnitude PDFs for the photometric sample and the specphot sample. The PDFs are evaluated at the $r'$ mag of each source, with the individual weights equal to the photometric sample PDF divided by the specphot PDF. These weights emphasize faint sources, which are underrepresented in the SDSS specphot sample. As galaxies outnumber stars by a ratio of $\sim$2:1 in the SDSS photometric observations, a weighted random selection of 200,000 galaxies and 100,000 stars from the SDSS specphot sample is made. We hereafter refer to these 300,000 sources as the SDSS field subset.

The ROC-like curve for the SDSS field subset is shown via the solid light-green line in Figure~\ref{fig:SDSS_cuts}. Again, the solid star shows the location of the $f_\mathrm{PSF}/f_\mathrm{cmodel} = 0.875$ threshold. Adopting the SDSS classification for all photometric sources detected by SDSS would yield a $\mathrm{FPR} \approx 0.07$. Additionally, a requirement of $\mathrm{FPR} = 0.005$ over all SDSS photometric detections would yield a $\mathrm{TPR} \approx 0.04$, which is so small it is effectively useless for screening point sources from the transient-candidate stream. Figure~\ref{fig:SDSS_cuts} also shows the ROC-like curves for the SDSS field subset restricted to sources with $r' \le 20$, 21, and 22 mag via dashed lines from top to bottom. The dashed lines confirm the previous assertions that the fidelity of the SDSS photometric classifier degrades rapidly for $r' > 21 \; \mathrm{mag}$. Thus, we supplement the PTF point-source catalog with all SDSS photometric detections satisfying $r' \le 21 \; \mathrm{mag}$ and $f_\mathrm{PSF}/f_\mathrm{cmodel} > 0.9658$.\footnote{The online SDSS documentation states that sources with $m_\mathrm{PSF} - m_\mathrm{cmodel} < 0.145 \; \mathrm{mag}$ are classified as stars, which is equivalent to the $f_\mathrm{PSF}/f_\mathrm{cmodel} > 0.875$ threshold discussed here. In terms of magnitude differrence, the adopted point-source classification threshold corresponds to $m_\mathrm{PSF} - m_\mathrm{cmodel} < 0.037 \; \mathrm{mag}$.} For sources with $r' \le 21 \; \mathrm{mag}$ this corresponds to a $\mathrm{TPR} = 0.79$ at the desired $\mathrm{FPR} = 0.005$. Thus, relative to the PTF RF point-source catalog, SDSS provides a $\sim$12\% increase in the recovered point sources at the desired FPR.  

The difference between our selection of SDSS point sources and that of the SDSS pipeline is illustrated in the bottom panel of Figure~\ref{fig:SDSS_cuts}. The density of the SDSS specphot sample is shown in the $f_\mathrm{PSF}/f_\mathrm{cmodel}-r'$ plane. There is a clear delineation between sources with $f_\mathrm{PSF}/f_\mathrm{cmodel} \approx 1$ and those with a larger model flux than PSF flux. The horizontal dashed line represents the SDSS classification threshold, while we only classify those sources enclosed by the solid pink line as point sources. Our cut is more restrictive, and produces a factor of $\sim$3 decrease in the number of galaxies erroneously classified as point sources.

\section{Summary and Conclusions}\label{sec:conclusions}

We have presented a method for the automated classification of stars and galaxies in PTF imaging data. The classifier utilizes the random forest algorithm and is trained using $> 3\times{10}^6$ PTF sources with SDSS spectra. A non-negligible fraction of point sources in the training set ($\sim$2\%) are photometric blends, targeted due to the SDSS bias to observe galaxies, especially LRGs. These blends, along with compact galaxies and low redshift quasars, were removed from the training set to improve the overall performance of the classifier. Features were selected from \texttt{SExtractor} shape and brightness measurements and the model tuning parameters were optimized via cross validation on the training set. 

We showed that the final PTF RF model outperformed the \texttt{SExtractor} classifications, with an overall improvement of $\sim$4\% on the \textit{photometrically clean} test set, and a more impressive $\sim$19\% improvement for sources with $r' \ge 21$ mag. Within the SDSS footprint, which covers roughly half of the total PTF imaging area, the SDSS pipeline provides better classifications than the PTF RF model due to the superior seeing in SDSS images. The PTF RF model produces near perfect separation of stars and galaxies down to $\sim$19 mag. Tests on a random selection of field stars show that the classification accuracy remains above 80\% down to $\sim$21 mag. To generate our final PTF point-source catalog we apply a conservative classification cut, designed to produce an $\mathrm{FPR} = 0.005$. Ultimately, only sources classified as stars in $\ge$725 of the 750 RF trees, corresponding to a RF relative ranking of $>$0.966, are included. 

In sum, there are $\sim$$1.70\times{10}^8$ sources in the point source catalog, of which only $\sim$$10^6$ are expected to be galaxies. Following a \textit{non-exhaustive} search for transients missed by the NERSC catalog, we identify two transients that would have been detected using the PTF RF catalog. This search, which only covered the last two months of 2015, provides definitive evidence that the PTF RF catalog enables new discoveries. We have additionally developed a new method to select SDSS point sources with an $\mathrm{FPR} = 0.005$, which we use to supplement the PTF RF point-source catalog within the SDSS imaging footprint. The inclusion of these SDSS sources increases the number of point-source detections by $\sim$12\%. The catalog has been incorporated into the various PTF transient discovery pipelines, and candidates associated with point sources are now automatically rejected and removed from the stream. 

Despite the large number of sources in the PTF point-source catalog, our conservative cut on RF relative ranking means that nearly $\sim{8} \times 10^7$ point sources are currently excluded from the catalog (more if one includes the sources fainter than 21 mag). Moving forward, especially with an eye towards ZTF, there are several potential improvements that could be made to improve the fidelity of the model, particularly at faint magnitudes. 

PTF, which has been running since 2009, uses \texttt{SExtractor} version 2.8.6 in the IPAC imaging pipelines. Recent versions of \texttt{SExtractor} (e.g., v2.19.5) include a new parameter \texttt{SPREAD\_MODEL} \citep{Desai12}, which acts as a discriminant between the best fit PSF model and an exponential model. Initial tests with \texttt{SPREAD\_MODEL} show that it is useful for separating stars and galaxies \citep{Soumagnac15}. The inclusion of \texttt{SPREAD\_MODEL} in a ZTF star-galaxy model will yield improvements relative to the PTF RF model.

Additional improvements can be had via deeper co-adds, which will make it easier to detect extended emission from sources with $r' \gtrsim 22$ mag. ZTF surveys the sky at a rate that is $\sim$$15\times$ faster than PTF, which will enable the creation of deep reference images faster than is currently possible. For instance, $\sim$38\% of the sources in the training set were taken from co-adds of only 5 images, while $\sim$57\% are from co-adds of $\le 10$ images. These references have a depth similar to SDSS, while the co-adds of 50 images detect sources as faint as $R_\mathrm{PTF} \gtrsim 23.5 \; \mathrm{mag}$. For ZTF deeper reference images will be generated to construct a point-source catalog with higher fidelity. 

Finally, altogether superior modeling of the sources at the image level could improve the separation of stars and galaxies in PTF and ZTF data. This could include techniques as familiar as simple PSF fitting with \texttt{DAOphot} \citep{stetson1987}, to more advanced solutions that construct probabilistic models of the data, such as \texttt{the Tractor}\footnote{See http://thetractor.org/.} \citep{Lang16}.

The detection and characterization of fast transients in the coming years will be as much about software development as it is about improvements in instrumentation. While events that evolve and disappear on timescales $\lesssim$ 24 hr have already been discovered (e.g., \citealt{Cenko13}), future $\sim$real-time classifications of these rarities will require swift automated decisions. The optimal allocation of expensive follow-up resources requires the best possible rejection of false positives. One step in that direction is to identify as many faint stars as possible, as we have done here for PTF observations. While the primary motivation for constructing the PTF RF catalog is to better enable the search for fast transients, these efforts ultimately improve the search for all extragalactic transients.

We are now firmly in the age of GW detections \citep{Abbott16}, and the identification of an electromagnetic counterpart to a GW event stands out as one of the most challenging problems in astrophysics in the coming years. The search for such counterparts will monopolize the use of wide-field telescopes across the globe (e.g., \citealt{Abbott16a}). Without some means to significantly reduce the haystacks, however, the search for these needles will be hopeless. Minimizing the stages at which human inspection and intervention are required, by actively reducing the number of false positive candidates, will improve our chances of one day catching the elusive transients associated with GW events. 

\acknowledgments 

This project started as part of an undergraduate research project at the California Institute of Technology. We thank T.\ Prince for funding MMK during the summer of 2015.

We are extremely greatful to A.\ Thakar, and the entire SDSS CasJobs Helpdesk for assistance in performing the large crossmatch between PTF and SDSS spectroscopic sources. Without their assistance this study would not have been possible. Without the patience and aid of R.~Lupton we would not have been able to recreate SDSS photometric classifier. We are in debt to M.~M.~Kasliwal, who endured countless conversations on the appropriate threshold for point-source classification. With gratitude, we salute S.~B.~Cenko for useful suggestions on the comparison of the NERSC catalog and the PTF RF catalog. Finally, we thank the anonymous referee for suggestions that improved this manuscript.

AAM acknowledges support for this work by NASA from a
Hubble Fellowship grant: HST-HF-51325.01, awarded by STScI,
operated by AURA, Inc., for NASA, under contract NAS 5-26555. 
Part of the research was carried out at the Jet Propulsion 
Laboratory, California Institute of Technology, under a contract
with NASA. 

Funding for SDSS-III has been provided by the Alfred P. Sloan Foundation, the Participating Institutions, the National Science Foundation, and the U.S. Department of Energy Office of Science. The SDSS-III web site is http://www.sdss3.org/.

\textit{ Facilities:} 
\facility{Sloan}, \facility{PO:1.2m}

\textcopyright 2016. All rights reserved.

\end{document}